\pdfoutput=1

\documentclass[11pt,twoside,a4paper,cmspaper,final,collab]{cms-tdr}
\def\svnVersion{183709}\def\cmsCernNo{2012-324}\def\cmsCernDate{\today}\def\cmsMessage{Submitted to Physical Review D}

\begin{document}\cmsNoteHeader{TOP-11-010}

\hyphenation{had-ron-i-za-tion}
\hyphenation{cal-or-i-me-ter}
\hyphenation{de-vices}

\RCS$Revision: 174467 $
\RCS$HeadURL: svn+ssh://svn.cern.ch/reps/tdr2/papers/TOP-11-010/trunk/TOP-11-010.tex $
\RCS$Id: TOP-11-010.tex 174467 2013-02-28 16:14:33Z alverson $
\ifthenelse{\boolean{cms@external}}{%
\newcommand{\scotchrule[1]}{\centering\begin{ruledtabular}\begin{tabular}{#1}}
\newcommand{\donescotchrule}{\end{tabular}\end{ruledtabular}}
}{
\newcommand{\scotchrule[1]}{\centering\begin{tabular}{#1}\hline}
\newcommand{\donescotchrule}{\hline\end{tabular}}
}
\providecommand{\CLs}{\ensuremath{\mathrm{CL}_\mathrm{s}}\xspace}
\newlength\cmsFigWidth\setlength{\cmsFigWidth}{0.3\textwidth}
\cmsNoteHeader{TOP-11-010} 
\title{\texorpdfstring{Search for $\cPZpr$ resonances decaying to $\ttbar$ in dilepton+jets final states in pp collisions at $\sqrt{s} = 7$\TeV}{Search for Z' resonances decaying to ttbar in dilepton+jets final states in pp collisions at sqrt(s) = 7 TeV}}

\date{\today}

\abstract{
A search for resonances decaying to top quark-antiquark pairs is performed using
a dilepton+jets data sample recorded by the CMS experiment at the LHC
in $\Pp\Pp$ collisions
at $\sqrt{s} = 7$\TeV corresponding to an integrated luminosity of
5.0\fbinv. No significant
deviations from the standard model background are
observed. Upper limits are presented for the
production cross section times branching fraction of top quark-antiquark resonances
for masses from 750 to 3000\GeV.
In particular, the existence of a leptophobic topcolor particle $\cPZpr$
is excluded at the $95\%$ confidence level for
resonance masses $M_{\cPZpr} < 1.3$\TeV for $\Gamma_{\cPZpr} =
0.012M_{\cPZpr}$, and $M_{\cPZpr} < 1.9$\TeV for $\Gamma_{\cPZpr} =
0.10M_{\cPZpr}$.
}

\hypersetup{%
pdfauthor={CMS Collaboration},%
pdftitle={Search for Z' resonances decaying to ttbar in dilepton+jets final states in pp collisions at sqrt(s) = 7 TeV},%
pdfsubject={Z' to ttbar},%
pdfkeywords={CMS, physics}}

\maketitle 

\section{Introduction
\label{sec:introduction}}
Electroweak symmetry breaking is a cornerstone for the understanding
of particle physics. However, despite the spectacular phenomenological
success of the standard model (SM), and the recent observation of a new
boson at the Large Hadron Collider (LHC)~\cite{ATLASHiggs,CMSHiggs}, the
precise mechanism of electroweak symmetry breaking remains unknown. Various
new models have been proposed to explain this mechanism. One such class of
models, topcolor-assisted technicolor
 (TC2)~\cite{laneref31,technicolor,laneref32}, provides a dynamical
explanation for electroweak symmetry breaking and flavor symmetry breaking,
giving masses to the weak gauge bosons and fermions. Under
one of the scenarios of TC2, a heavy boson $\cPZpr$ is predicted with
preferential couplings to the third quark generation and with no
significant couplings to the leptons (``leptophobic'').

Direct searches for massive resonances that
decay preferentially to top quark-antiquark ($\ttbar$) pairs are
currently feasible only at hadron colliders. Experiments seek to
observe an excess beyond that predicted by the SM, typically in the
distribution of the invariant mass of the $\ttbar$ decay products. Searches in
$\Pp\Pap$ collisions at the Tevatron and the early searches in $\Pp\Pp$
collisions at the LHC by the ATLAS experiment
have excluded a narrow-width, leptophobic $\cPZpr$ with a mass lower than
900\GeV~\cite{cdfZprime,d0Zprime,atlasZprime}. The
searches by the Compact Muon Solenoid (CMS) experiment at the LHC have
excluded a narrow-width,
leptophobic $\cPZpr$ in the mass range 1.3--1.5\TeV and in a narrow
window around 1\TeV~\cite{CMShadronicZprime,CMSleptonicZprime}.
This paper describes a search for a $\cPZpr\rightarrow\ttbar$ resonance in
$\Pp\Pp$ collisions in the $2\ell + 2\nu+\text{jets}$ final state, where
$\ell$ is an electron ($\Pe$) or a muon ($\mu$). This is the
first search for topcolor leptophobic $\cPZpr$ in final states
involving two leptons. The data sample corresponds to a total
integrated luminosity of 5.0\fbinv~\cite{lumi}
at $\sqrt{s} = 7$\TeV collected by the CMS detector in 2011.

\section{The CMS detector
\label{sec:cmsdetector}}
The central feature of the CMS apparatus is a superconducting
solenoid, 13 m in length and 6\unit{m} in diameter, which provides an
axial magnetic field of 3.8\unit{T}. The bore of the solenoid is
surrounded by various particle detection systems. Charged particle
trajectories are measured by a silicon pixel and strip
tracker, covering $0 < \phi \leq 2\pi$ in azimuth and $|\eta| < 2.5$, where the
pseudorapidity $\eta$ is defined as $\eta = - \ln[\tan(\theta/2)]$, and $\theta$ is the
polar angle of the trajectory of the particle with respect
to the counterclockwise-beam direction. A crystal electromagnetic calorimeter
and a brass/scintillator hadronic calorimeter surround the tracking
volume. The calorimeter provides
high-resolution energy measurement of electrons.
Muons are measured in gas-ionization detectors embedded in the
steel flux return yoke outside the solenoid. The detector is nearly
hermetic, which facilitates the measurement of energy balance in the
plane transverse to the beam direction. A two-tier trigger system selects
the most interesting $\Pp\Pp$ collision events for use in physics
analysis. A more detailed description of the CMS detector can be
found in Ref.~\cite{cmsdet}.

\section{Event reconstruction
\label{sec:selection}}
In the $\cPZpr \rightarrow \ttbar$ search, a $\ttbar$ decay topology is used
where each top quark decays to a $\PW$ boson and a $\cPqb$ quark, and
subsequently each $\PW$ boson decays into a lepton and a neutrino.
The signature for such an
event is two oppositely charged, isolated leptons with high
transverse momenta ($\pt$), large momentum imbalance due to
two undetected neutrinos, and
at least two jets. Events are required to pass a trigger requiring at
least two high-$\pt$ isolated
leptons, and are separated into three channels based
on lepton flavor: $\Pe\Pe$, $\mu\mu$,  and $\Pe\mu$. The principal
sources of background are SM $\ttbar$,
$\cPZ/\gamma^*\rightarrow\ell\overline\ell$ (Drell--Yan, DY), single-top-quark, and
diboson ($\PW\PW$, $\PW\cPZ$, and $\cPZ\cPZ$) production. Other
minor contributions are from $\PW\rightarrow \ell\nu$ and multijet
production. Electrons,
muons, jets, and the momentum imbalance are reconstructed using a
particle-flow algorithm~\cite{pflow-cms}. The negative of the vector
sum of the momenta of all reconstructed particles
in the plane transverse to the beams is the missing transverse
momentum~\cite{metreco}, whose magnitude is called missing
transverse energy ($\ETslash$). The identification criteria
of each object and additional selections are chosen to reduce all
backgrounds other than SM $\ttbar$ production.

Electron candidates are reconstructed from clusters of energy deposits
in the electromagnetic calorimeter, which are then matched to hits in the
silicon tracker. Electron identification is based on shower-shape and
track-cluster matching variables~\cite{electronreco}. Electrons are
required to have
$\pt > 20$\GeV and $|\eta| < 2.5$, and are excluded if they are in
the transition region between the barrel and endcap calorimeters,
$1.4442 < |\eta| < 1.5560$, because their reconstruction in
this region is degraded due to additional material there. The electron
track must pass within 0.04\unit{cm} of the primary vertex in the plane
transverse to the beam.
Additionally, electrons coming from photon conversions in the detector
material are rejected if there are missing hits in the inner tracker
layers or if there is another close track with opposite charge and
with a similar polar angle.

Muons are reconstructed using the information from the muon
detectors and the silicon tracker~\cite{muonreco}.
The reconstructed muon track must be within 0.02\unit{cm} of the primary vertex in the
plane transverse to the beam. Muons are
required to have $\pt > 20$\GeV and $|\eta| < 2.4$.

To remove leptons arising from decays of hadrons
immersed in jets, the electrons and muons are required to be
isolated. The
isolation requirement is based on the ratio of the total transverse
energy observed from all particles in a cone of size
$R = \sqrt{(\Delta\phi)^2 + (\Delta\eta)^2} < 0.3$ centered on the
direction of the lepton to the transverse momentum of the lepton.
This quantity must be less than 0.17 for electrons, and less than 0.20 for
muons.

In order to reduce the background from low-mass dilepton resonances,
events are rejected if the dilepton invariant mass
$M_{\ell\ell} < 12$\GeV. To suppress the contribution from $\cPZ$-boson
production, a veto on events with $76 < M_{\ell\ell} < 106$\GeV is
applied in the $\Pe\Pe$ and $\mu\mu$ channels.

Events are required to contain at least two jets, reconstructed
using an anti-\kt clustering algorithm with a distance parameter
of 0.5~\cite{jetkt}. Corrections are applied to account for the dependence of
the detector response to jets as a function of $\eta$ and $\pt$ and the
effect of pileup (multiple $\Pp\Pp$ collisions)~\cite{JETJINST}.
The corrections are based on in-situ calibration using dijet and
$\gamma$/$\cPZ$+jet samples. In the $\pt$ region above $\sim$1\TeV, where
the statistics of the calibration samples become insufficient,
the jet energy scale is constrained using the single particle response
from test beam data~\cite{JETJINST}.
All corrections are propagated to recalculate the missing transverse energy.
The jets are required to have $\pt > 30$\GeV and $|\eta| < 2.5$.
Additionally, at least one of the jets is required to be tagged as
a $\cPqb$ jet based on the identification of a secondary
vertex~\cite{CMS-PAS-BTV-11-004}.

Finally, in order to further reduce the DY and multijet backgrounds in the $\Pe\Pe$
channel, and the DY contribution in the $\mu\mu$
channel, a requirement of $\ETslash > 30$\GeV is applied in these channels.
The DY and multijet backgrounds are negligible in the $\Pe\mu$ channel.

\section{Signal and background modeling
\label{sec:model}}
The signal efficiency and background rejection of the selection outlined
above are determined from simulation studies augmented where necessary
by corrections based on control samples in data.
The resonance signal $\cPZpr \rightarrow \ttbar$ is modeled using the
{\MADGRAPH}~5.1.1~\cite{madgraph5} Monte Carlo (MC) event generator, with
the top-quark mass ($M_\cPqt$) set to 172.5\GeV and CTEQ6L~\cite{cteq6L} parton distribution
functions. Samples are generated with resonance masses
between 750--3000\GeV, and for two resonance-width scenarios:
narrow ($\Gamma_{\cPZpr} = 0.012M_{\cPZpr}$) and wide
($\Gamma_{\cPZpr} = 0.10M_{\cPZpr}$). To calculate the expected number of
signal events, we use cross sections for a leptophobic topcolor
$\cPZpr$~\cite{jainharris}. A scale factor of 1.3 is used to account for
next-to-leading-order (NLO) corrections~\cite{ZprimeKfactor}.

The background events from SM $\ttbar$, DY, and
$\PW\rightarrow \ell\nu$ are generated
using \MADGRAPH~5.1.1. Diboson events are generated
using \PYTHIA~6.424~\cite{pythia}, and single-top-quark events
are generated using
\POWHEG~1.0~r1380~\cite{powheg2009,powheg2010,powheg2011}.
The same set of parton distribution functions are used for each
process as for the
resonance signal $\cPZpr$. The \MADGRAPH and
\POWHEG events are processed through \PYTHIA in order to
add the initial- and final-state radiation and showering, together with
the production of the underlying event~\cite{pythiatune}.
To estimate the expected number of
background events, the background samples are normalized to the
theoretical cross sections shown in Table~\ref{tab:mcxsec}.
\begin{table}[htbH]
\topcaption{Theoretical cross sections, including higher-order corrections,
  for the SM backgrounds~\cite{SMxsec-tt,SMxsec-wjets,SMxsec-tchannel,SMxsec-schannel,SMxsec-tWchannel}.
\label{tab:mcxsec}}
\scotchrule[lc]
Background & Cross section [pb] \\
\hline
$\ttbar$
&  160 \\ 
DY ($M_{\ell\ell} > 10$\GeV)
& 15000 \\ 
${\PW\PW}\rightarrow 2{\ell}2\nu$
& 4.5 \\  
${\PW\cPZ}\rightarrow 3{\ell}\nu$
& 0.61 \\
${\cPZ\cPZ}$ (inclusive)
& 7.4 \\
${\PW}\rightarrow \ell\nu$
& 31000 \\ 
Single top
& 85 \\  
\donescotchrule
\end{table}
All MC events are processed
through a simulation of the CMS detector based on
\GEANTfour~\cite{geant4}, and are overlaid with events
from minimum-bias interaction to account for pileup effects at high
instantaneous luminosity. An additional set of corrections is applied
to account for data-MC differences. These include reweighting MC events
to match the overlaid pileup distribution to that inferred from the data, and
applying a scale factor of 0.95 per tagged $\cPqb$ jet~\cite{CMS-PAS-BTV-11-004} to
account for the observed difference in $\cPqb$-tagging efficiency between data and MC.
The \cPqb-tagging efficiency is measured using $\ttbar$ and muon+jets
events~\cite{CMS-PAS-BTV-11-004}, with the uncertainty on the data/MC scale factor
amounting to 10--20\%.

The simulation of the DY background does
not adequately reproduce the production rate, especially in the
presence of missing transverse energy. The overall normalization for
the DY process is therefore obtained using data in the
DY-enriched region of $76 < M_{\ell\ell} < 106$\GeV in
the $\Pe\Pe$ and $\mu\mu$ channels. This region is excluded from the
dataset used for the main analysis and thus provides an
independent DY control sample.
In the $\Pe\mu$ channel, the DY-enriched region is
between $40 < M_{\ell\ell} < 70$\GeV since the main contribution to DY in this channel
is from $\cPZ/\gamma^*\rightarrow\tau\bar\tau$ events. These events
have a peak at lower dilepton masses on account of the invisible decay products of
the $\tau$ lepton.
As the \cPZ-mass veto is not applied in the $\Pe\mu$ channel, the requirement of at
least two jets is modified to exactly one jet in order to ensure the exclusion
of this calibration sample from the main signal sample. The
normalization factors obtained are
$1.34 \pm 0.025$,
$1.24 \pm 0.021$, and
$1.20 \pm 0.052$
in the \Pe\Pe, $\mu\mu$, and $\Pe\mu$ channels, respectively.

The multijet background is estimated directly from data. This
background is from misidentified leptons or genuine leptons
from semi-leptonic decays of \cPqb/$\cPaqb$ or \cPqc/$\cPaqc$
quarks, which pass the isolation requirement. It is
determined from data by inverting the
isolation criteria for both leptons, and then
extrapolating that yield to the signal region. The extrapolation is
performed by multiplying the yield by a normalization factor
that accounts for the isolation efficiency obtained from like-sign
lepton events, defined as:
\begin{equation}
\label{eq:qcdisoeff}
	f_\mathrm{QCD} =   \frac{N_\text{data}^{\pm\pm\text{,isolated}} -
          N_\mathrm{MC}^{\pm\pm\text{,isolated}}}{N_\text{data}^{\pm\pm\text{,non-isolated}}},
\end{equation}
where $N ^{\pm\pm}$ represents the number of like-sign
lepton events with the isolation criteria on both leptons
either applied or inverted, and MC represents background
predictions for $\ttbar$, DY, diboson,
$\PW\rightarrow \ell\nu$, and single-top quark events.
The multijet estimate was cross-checked with an alternate method in a similar
analysis with the same final state, and good agreement was
observed~\cite{arxiv1211-2220}.

\begin{figure*}[hbt]
 \centering
  \includegraphics[width=\cmsFigWidth]{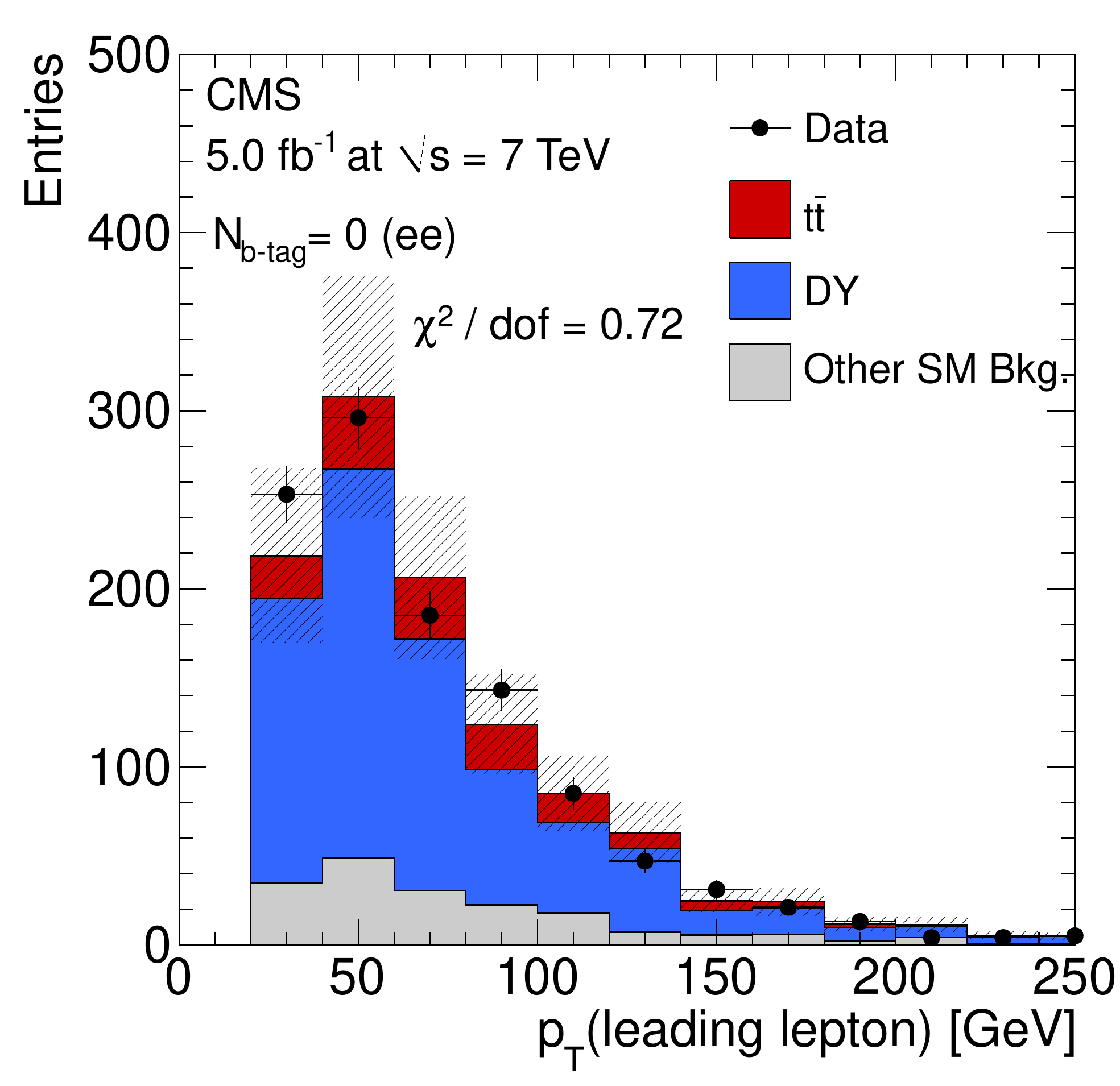}
  \includegraphics[width=\cmsFigWidth]{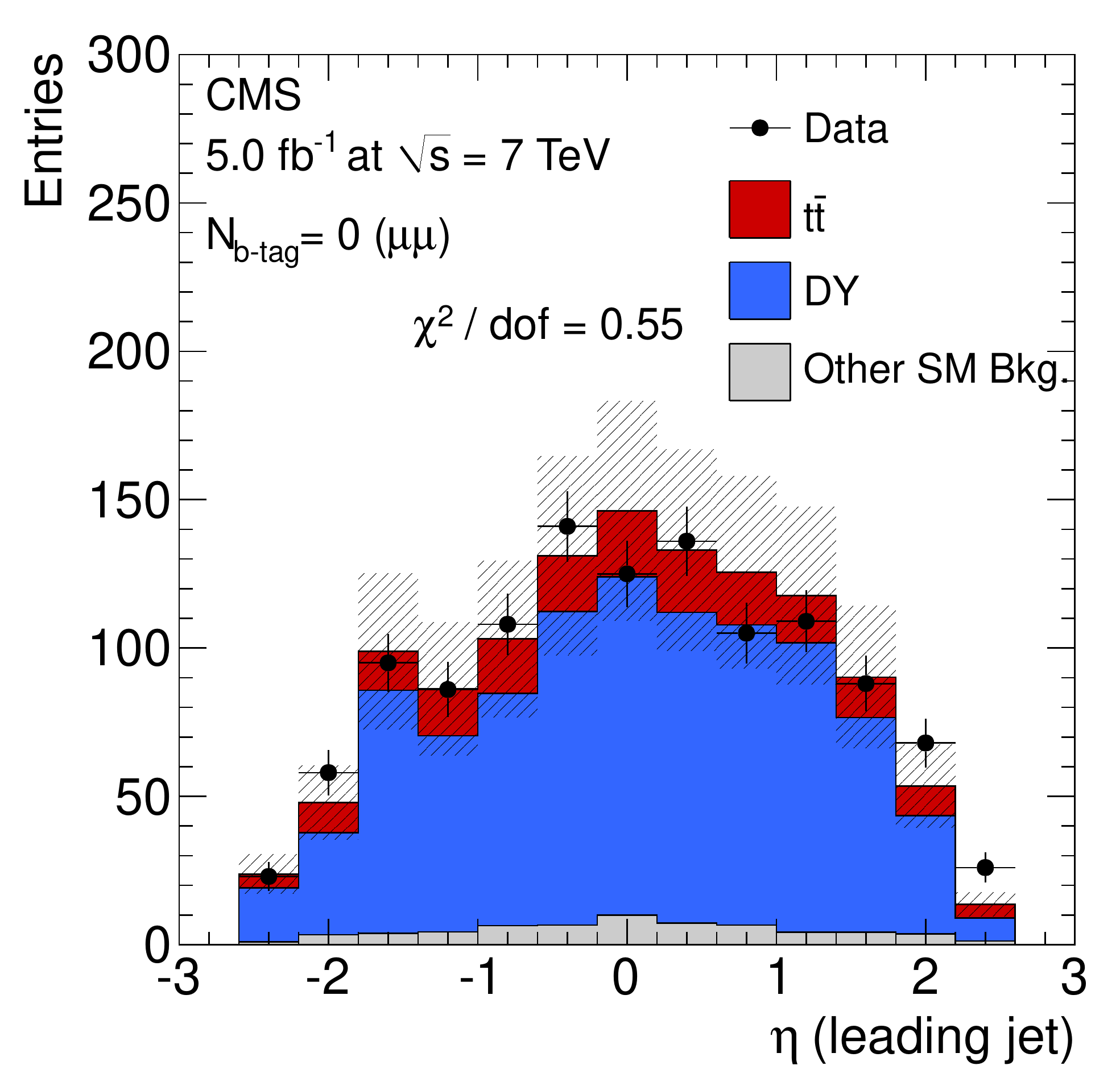}
  \includegraphics[width=\cmsFigWidth]{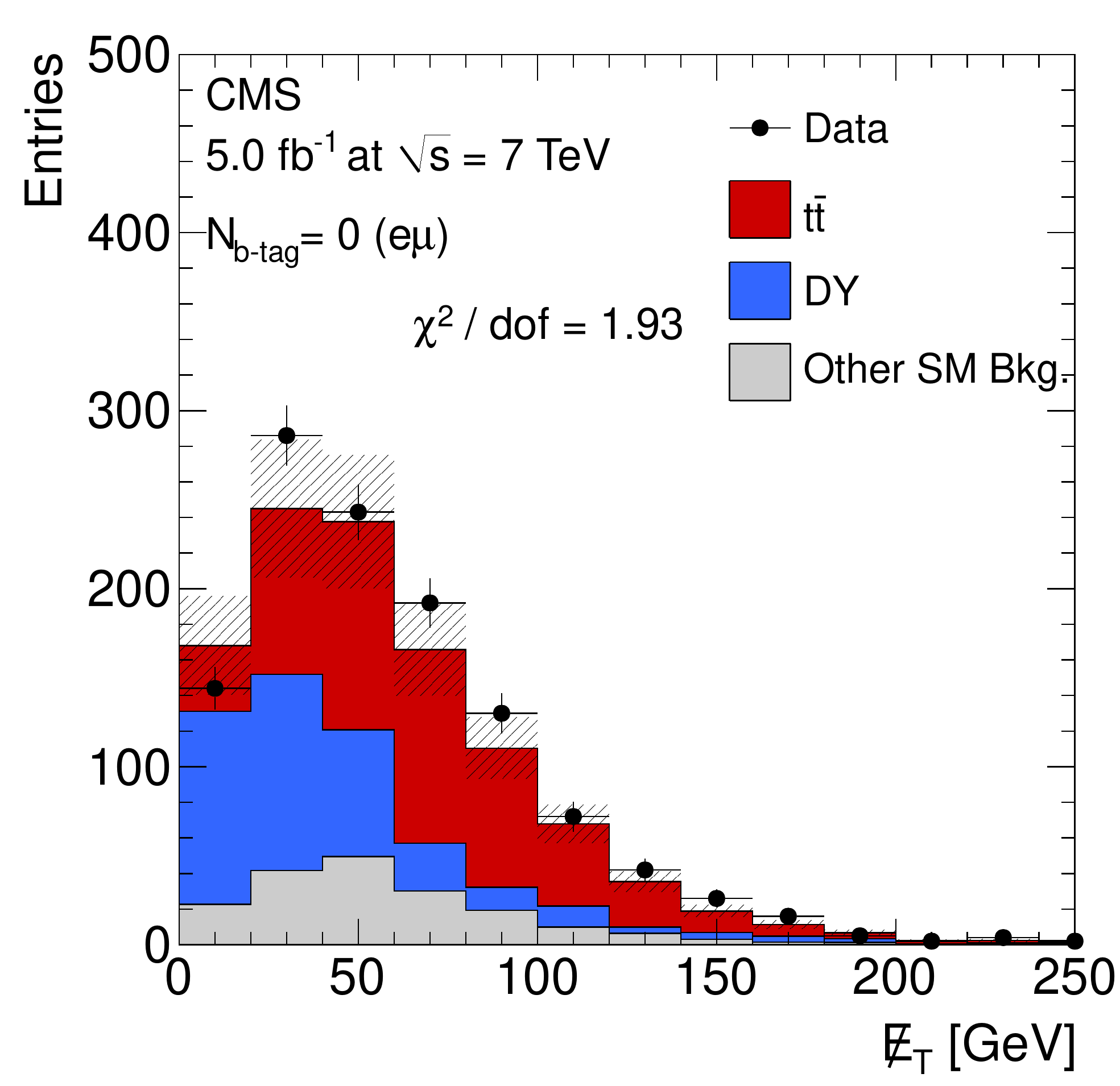}
 \caption{Distributions of the transverse momentum of the highest-\pt
electron in the $\Pe\Pe$ channel, the pseudorapidity of the highest-\pt
jet in the $\mu\mu$ channel, and the missing
transverse energy in the $\Pe\mu$ channel in the control sample (zero
\cPqb-tagged jets). The hatched region indicates systematic uncertainties on
the sum of SM backgrounds discussed in Section~\ref{sec:systematics}.
\label{fig:controlvars}}
\end{figure*}

In order to check the background modeling, a control sample is created by
requiring zero $\cPqb$-tagged jets. This sample is dominated
by non-$\ttbar$ backgrounds. Figure ~\ref{fig:controlvars} shows
representative distributions from this sample in the $\Pe\Pe$,
$\mu\mu$, and $\Pe\mu$ channels. The shape of the multijet background
distribution is derived directly from  the data using a control sample
of non-isolated leptons, in contrast to the other background shapes,
which are taken from simulation. Good agreement is observed between
data and background prediction in all three channels.

\section{Event yields
\label{sec:yield}}
The number of events for the expected SM backgrounds and
the observed data after all selections and corrections to account for
data-MC differences are listed in Table~\ref{tab:yield}. The uncertainties
shown for the backgrounds are from the systematic effects discussed in
Section~\ref{sec:systematics}. There is good agreement between data
and SM backgrounds in all three channels.
For comparison, the numbers of expected $\cPZpr$ signal events for
$M_{\cPZpr} =750$--3000\GeV are listed in Table~\ref{tab:yieldZprime}.
\begin{table*}[htbH]
\topcaption{Event yields for SM backgrounds and data in the
\Pe\Pe, $\mu\mu$, and $\Pe\mu$ channels.
\label{tab:yield}}
\scotchrule[lccc]
Sample  & \Pe\Pe & $\mu\mu$ & $\Pe\mu$ \\
\hline
$\ttbar$                     & 2210 $\pm$ 460 &  2550 $\pm$ 550 &  7300 $\pm$ 1500 \\
DY &  410 $\pm$ 130 &   420 $\pm$ 130 &   179 $\pm$   57 \\
Diboson                      &   11.5 $\pm$   1.5 &    15.4 $\pm$   2.0 &    32.3 $\pm$    4.2 \\
$\PW\rightarrow \ell\nu$       &   $17^{+24}_{-17}$  &     $0^{+3.4}_{-0}$ &  $26^{+37}_{-26}$   \\
Single top                &  106 $\pm$  15 &   121 $\pm$  18 &   343 $\pm$   50 \\
Multijets                 &   41.8 $\pm$   6.9 &    50 $\pm$  10 &   103 $\pm$   14 \\
\hline
Total background          & 2790 $\pm$ 510 & 3150 $\pm$ 600 & 8000 $\pm$ 1500 \\
\hline
Data       & 2690 & 3098 & 7704 \\
\donescotchrule
\end{table*}
\begin{table}[htbH]
\topcaption{Event yields for a leptophobic $\cPZpr$ in \Pe\Pe, $\mu\mu$, and $\Pe\mu$ channels.
\label{tab:yieldZprime}}
\scotchrule[lccc]
Sample  & \Pe\Pe & $\mu\mu$ & $\Pe\mu$ \\
\hline
\multicolumn{4}{c}{$\Gamma_{\cPZpr}/M_{\cPZpr} = 0.012$}\\
$\cPZpr$ (750\GeV)            &  67 &  79 & 200  \\
$\cPZpr$ (1000\GeV)           &  26 &  28 &  68  \\
$\cPZpr$ (1250\GeV)           &   8.2 &   9.8 &  22  \\
$\cPZpr$ (1500\GeV)           &   2.9 &   3.1 &   7.0  \\
$\cPZpr$ (2000\GeV)           &   0.3 &   0.4 &   0.8  \\
\hline
\multicolumn{4}{c}{$\Gamma_{\cPZpr}/M_{\cPZpr} = 0.10$} \\
$\cPZpr$ (1000\GeV)           &  180 & 200 & 480  \\
$\cPZpr$ (1500\GeV)           &   23 &  26 &  57  \\
$\cPZpr$ (2000\GeV)           &    2.9 &   2.9 &   7.0  \\
$\cPZpr$ (3000\GeV)           &    0.1 &   0.1 &   0.2  \\
\donescotchrule
\end{table}

\begin{figure*}[hbt]
 \centering
  \includegraphics[width=\cmsFigWidth]{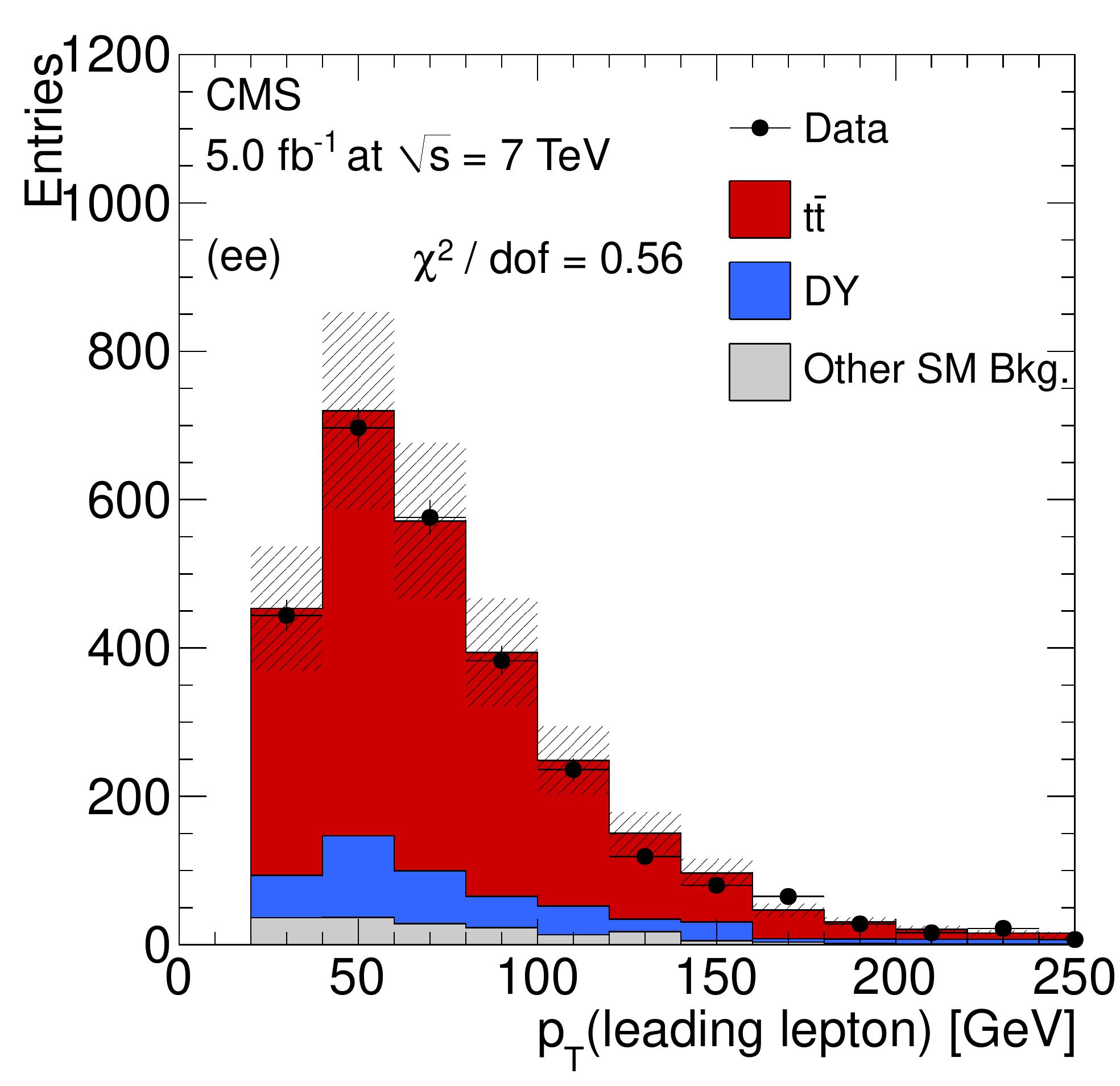}
  \includegraphics[width=\cmsFigWidth]{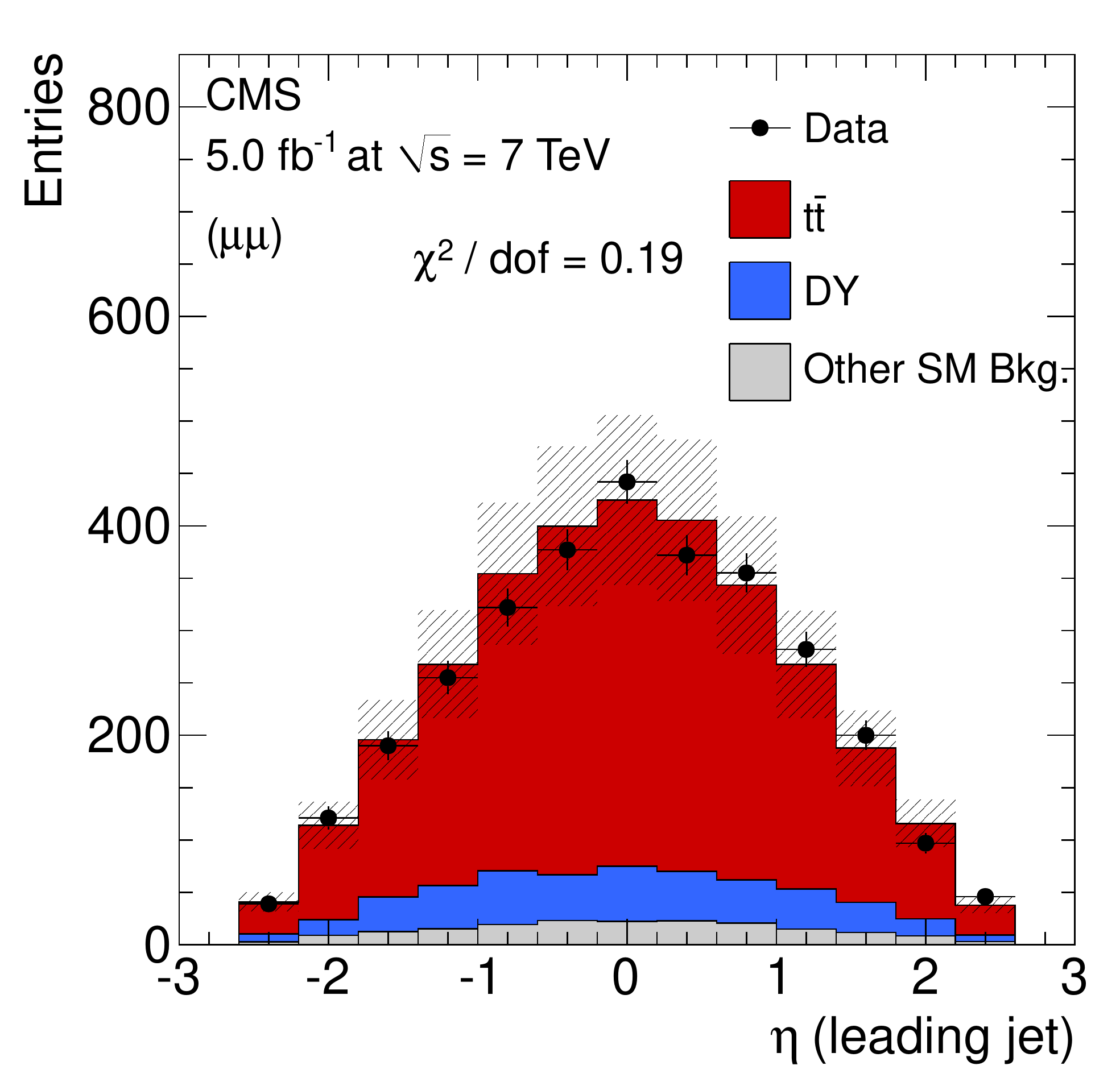}
  \includegraphics[width=\cmsFigWidth]{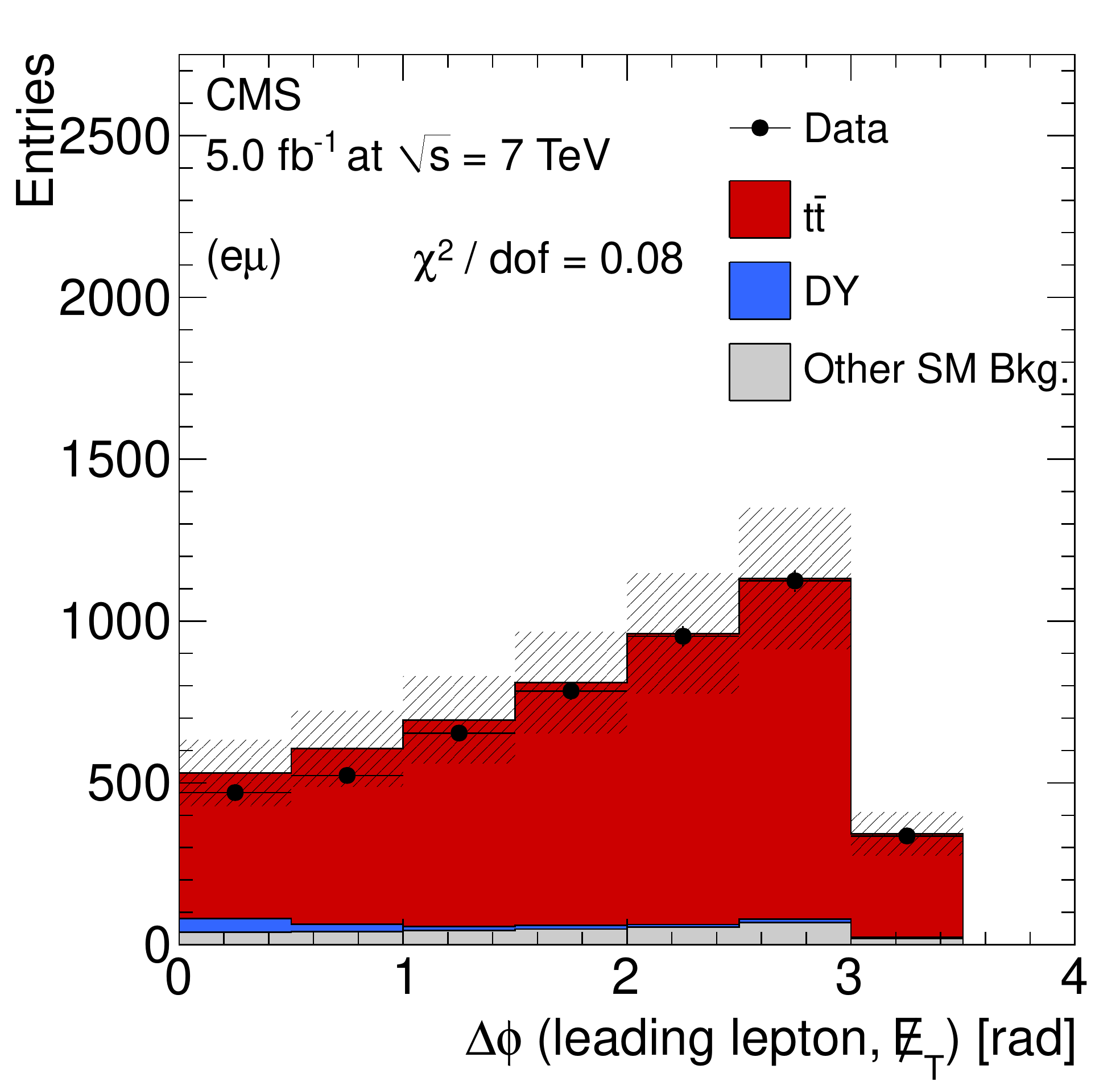} \\
  \includegraphics[width=\cmsFigWidth]{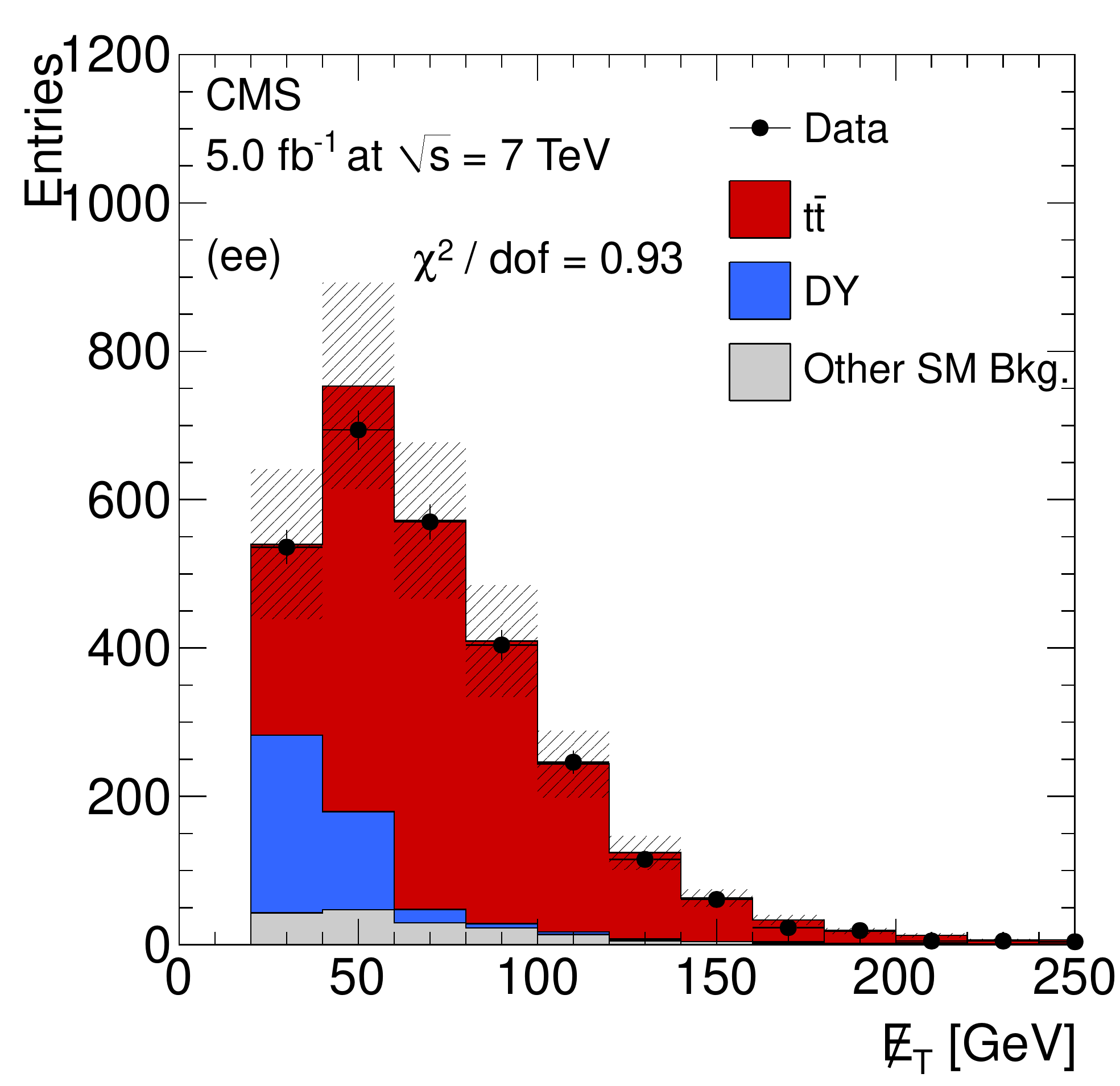}
  \includegraphics[width=\cmsFigWidth]{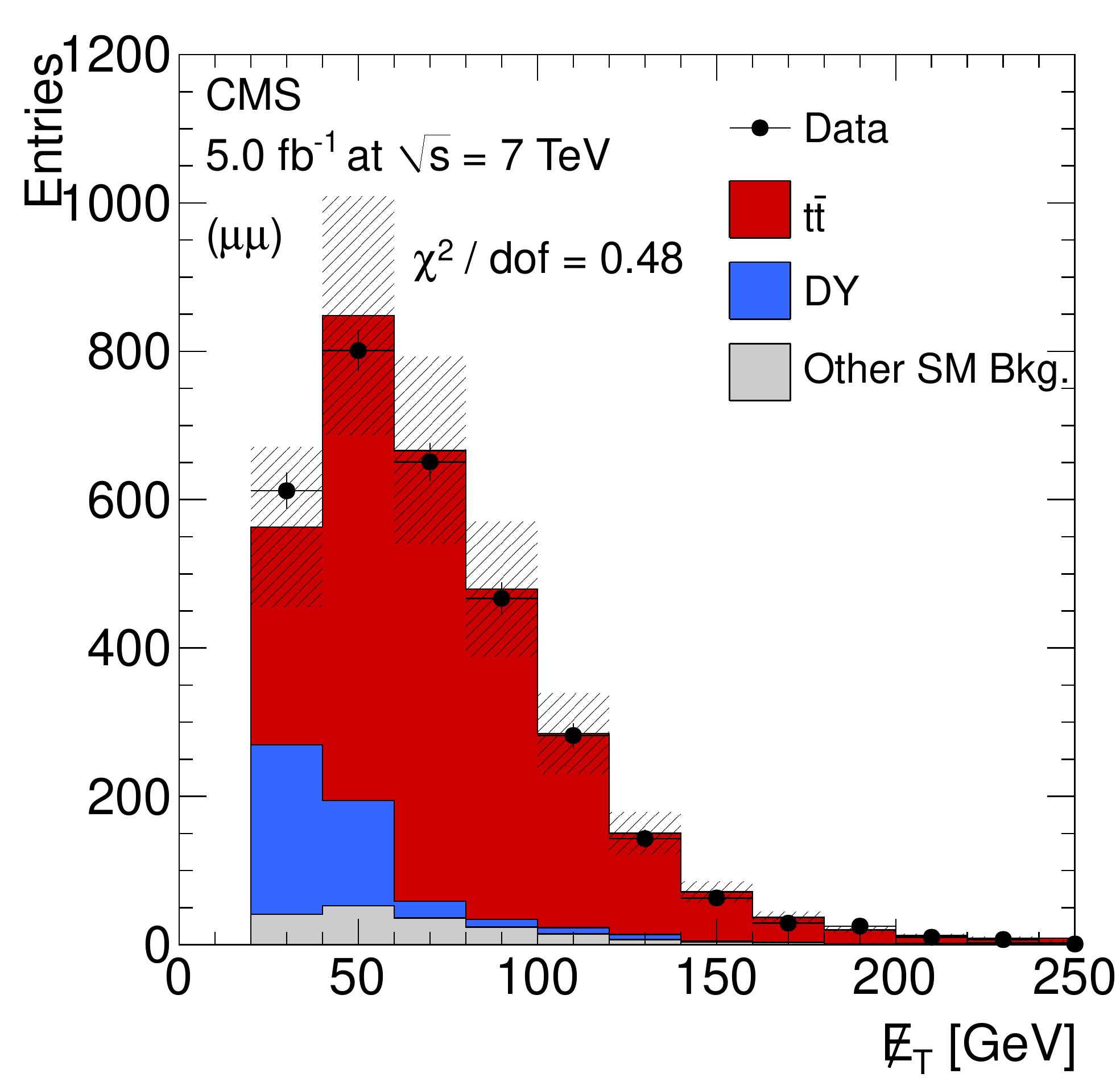}
  \includegraphics[width=\cmsFigWidth]{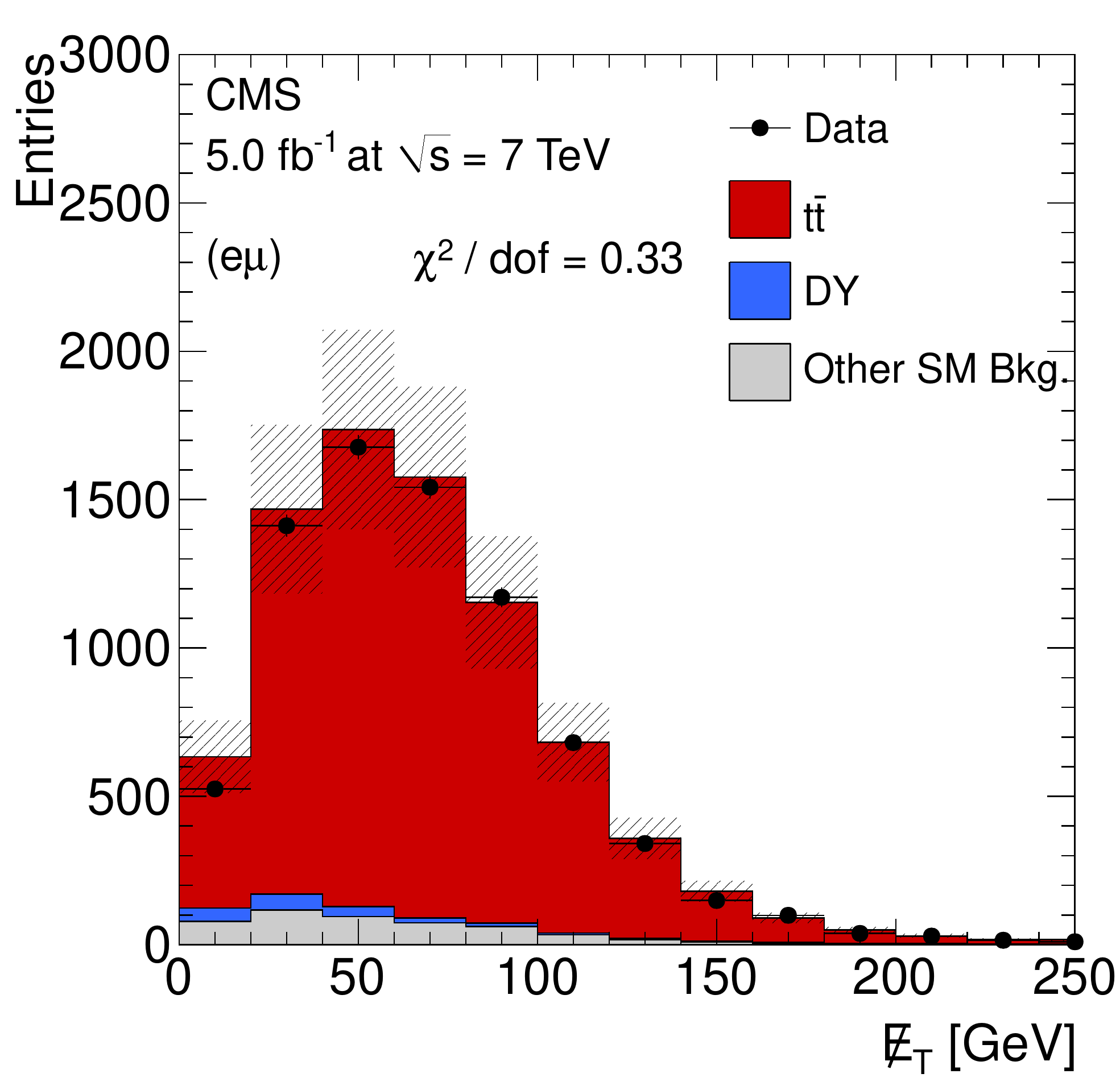}
 \caption{Distributions of the transverse momentum of the highest-\pt
electron in the $\Pe\Pe$ channel, the pseudorapidity of the highest-\pt
jet in the $\mu\mu$ channel, and $\Delta\phi$ between the missing
transverse momentum and highest-\pt lepton in the $\Pe\mu$ channel
(top row), and the missing transverse energy in the three channels
(bottom row). The hatched region indicates systematic uncertainties on
the sum of SM backgrounds.
\label{fig:kinemvars}}
\end{figure*}

Distributions of the transverse momentum of the highest-\pt
electron in the $\Pe\Pe$ channel, the pseudorapidity of the highest-\pt
jet in the $\mu\mu$ channel, and $\Delta\phi$ between the missing
transverse momentum and highest-\pt lepton in the $\Pe\mu$ channel are
shown in Fig.~\ref{fig:kinemvars}. Also shown are the distributions of
the missing transverse energy in the three channels.
There is good agreement between data
and the sum of all SM backgrounds. Similarly, a good agreement is seen in all
four-vector distributions (\pt, $\eta$, and $\phi$) of all
final-state objects in the three channels.

\begin{figure*}[hbt]
 \centering
  \includegraphics[width=\cmsFigWidth]{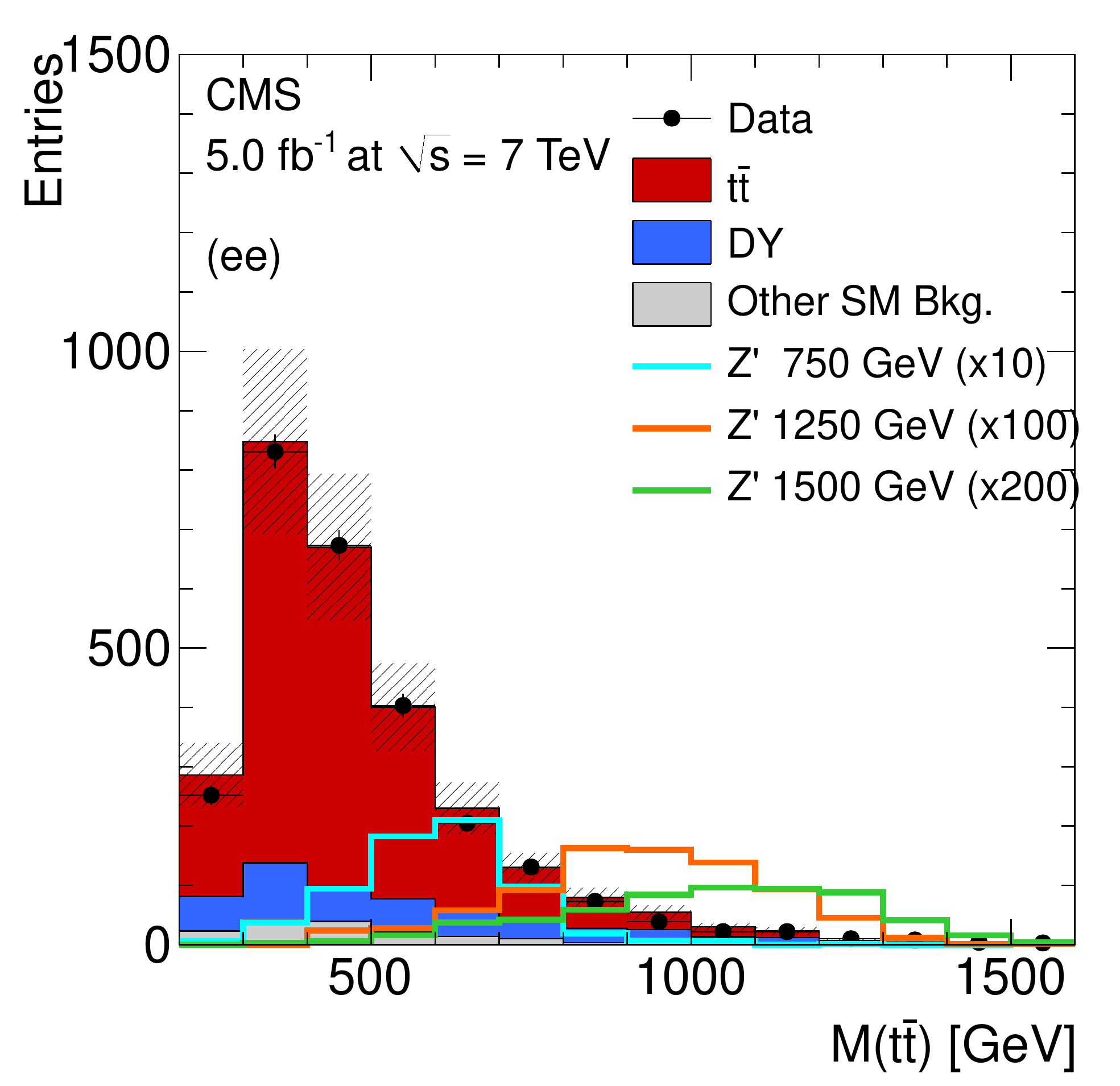}
  \includegraphics[width=\cmsFigWidth]{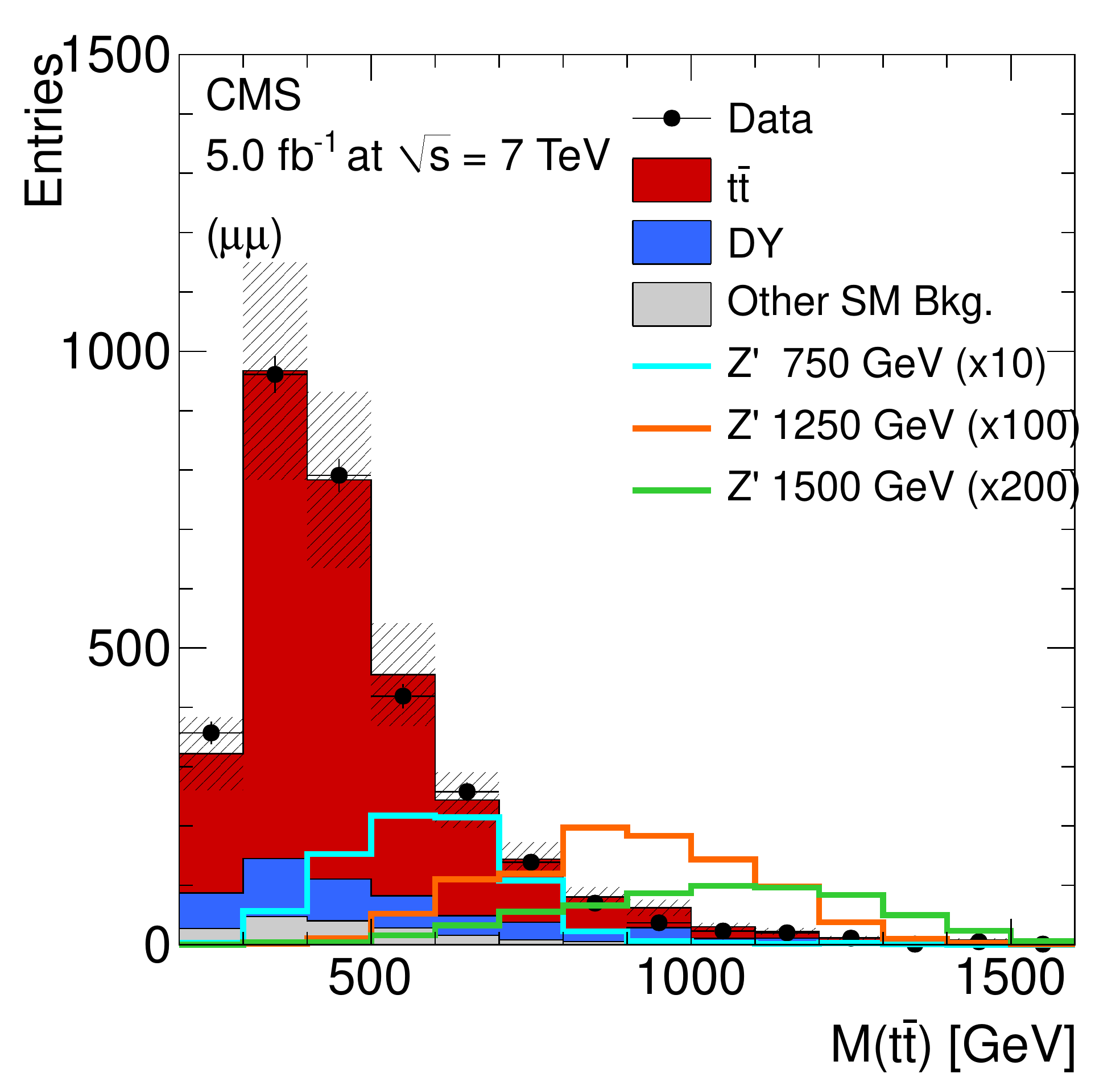}
  \includegraphics[width=\cmsFigWidth]{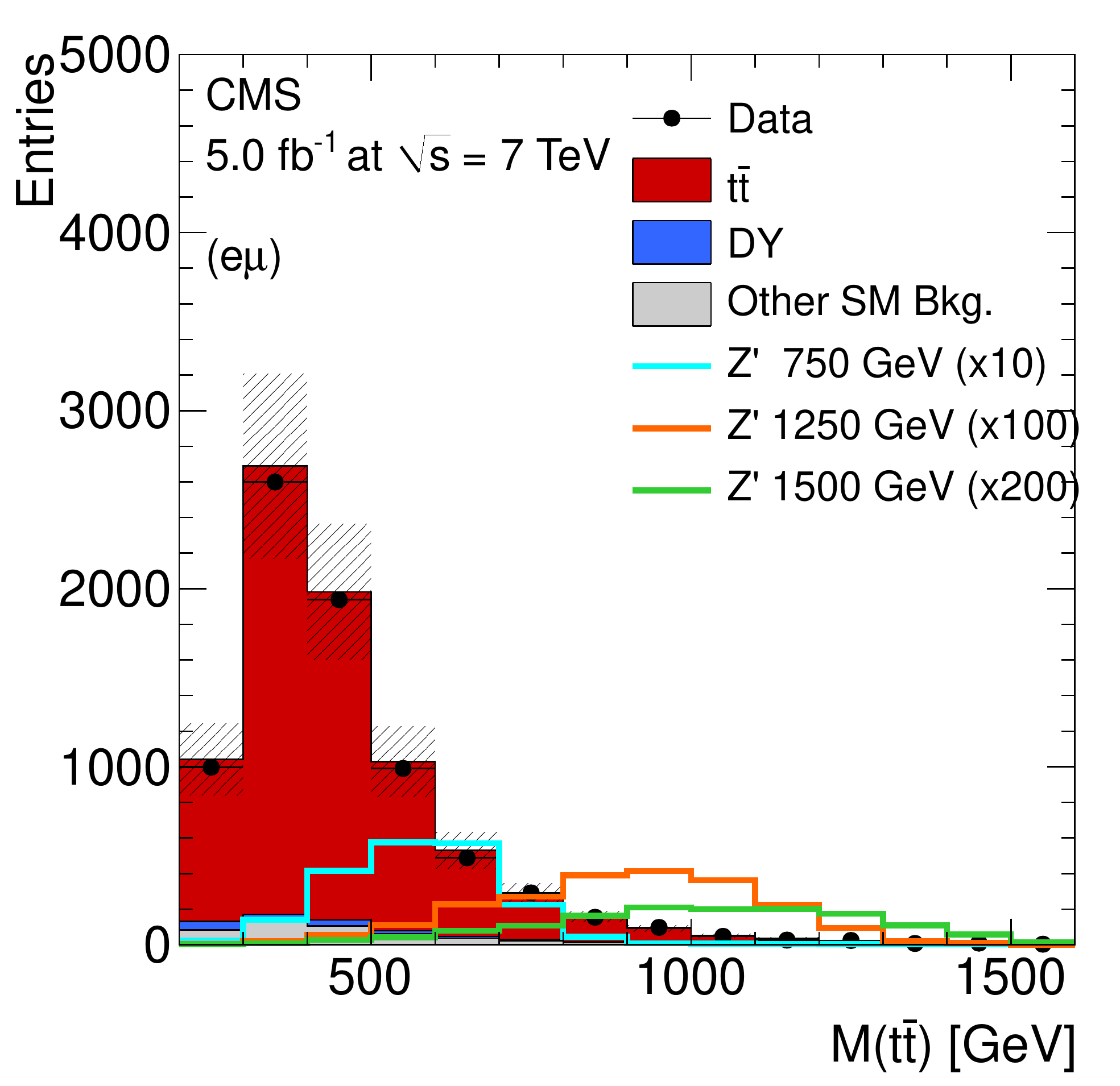}
 \caption{Distributions of the $\ttbar$ invariant mass for the \Pe\Pe,
   $\mu\mu$, and $\Pe\mu$ channels. The $p_z$ values
   for both neutrinos are set to zero. The
hatched region indicates systematic uncertainties on the sum of SM
backgrounds. The $\cPZpr$ signal corresponds to a resonance width of
$\Gamma_{\cPZpr} = 0.012M_{\cPZpr}$, and has been scaled up so as to be visible.
\label{fig:invmass}}
\end{figure*}

The $\ttbar$ invariant mass is constructed using the four-vectors of
the two leading leptons, the two leading jets, and the missing
transverse energy. The longitudinal momenta $p_z$ of the two
neutrinos in the final state cannot be measured
experimentally and are set to zero. The $\ttbar$ invariant mass
distributions for the data, the sum of all
backgrounds, and the $\cPZpr$ signals for the narrow-width scenario
($\Gamma_{\cPZpr} = 0.012M_{\cPZpr}$) are shown in Fig.~\ref{fig:invmass}
for the \Pe\Pe, $\mu\mu$, and $\Pe\mu$ channels. The
data are described well by the SM backgrounds, and there is no
statistically significant evidence for the presence of a $\cPZpr$ signal.

\section{Bayesian neural network analysis
\label{sec:bnn}}
A multivariate analysis, based on Bayesian neural
networks (BNN)~\cite{BNN}, has been carried out to provide a more
powerful discriminant between backgrounds and the $\cPZpr$ signal than
that based on invariant mass alone.
The discriminant lies in the interval [0,1]. It is constructed such that the
signal events tend to have a value closer to 1 than the background
events that peak instead closer to 0. To build this discriminant the
background is defined as the sum of all SM processes, and the signal
is set to $M_{\cPZpr} = 750$\GeV for narrow
width, and $M_{\cPZpr} = 1000$\GeV for wide width. The separation power
increases with hypothetical $\cPZpr$ mass.  Thus, using the $\cPZpr$
simulated sample with the lowest $\cPZpr$ mass for the signal when training the BNN
ensures good discrimination between signal and background even for the
higher $\cPZpr$ masses. A separate discriminant is constructed in each of
the three channels. As inputs to the training, the following 17
variables are used in each of the two scenarios, and in each channel:
\begin{itemize}
\item
$\pt$ and $\eta$ of the highest-\pt lepton,
\item
$\pt$, $\eta$, and $\Delta\phi$ of the second highest-\pt lepton,
\item
$\pt$, $\eta$, and $\Delta\phi$ of the highest-\pt jet,
\item
$\pt$, $\eta$, and $\Delta\phi$ of the second highest-\pt jet,
\item
$\ETslash$, and $\Delta\phi$ of the missing transverse momentum,
\item
$\pt$, $\eta$, and $\Delta\phi$ of the highest-\pt $\cPqb$-tagged jet,
and number $n_\cPqb$ of $\cPqb$-tagged jets,
\end{itemize}
where $\Delta\phi$ is the difference in azimuth between the object and the
highest-\pt lepton. All input variables are internally transformed by the BNN
to have a range of $[-1, 1]$. This set of input variables constitutes the full
array of four vectors of final-state objects that are measured in the
analysis, along with additional information about the $\cPqb$-tagged
jets. Using additional derived quantities such as the reconstructed
$\ttbar$ invariant mass as an input to the BNN does not improve the
performance of the BNN.

\begin{figure*}[hbt]
 \centering
  \includegraphics[width=\cmsFigWidth]{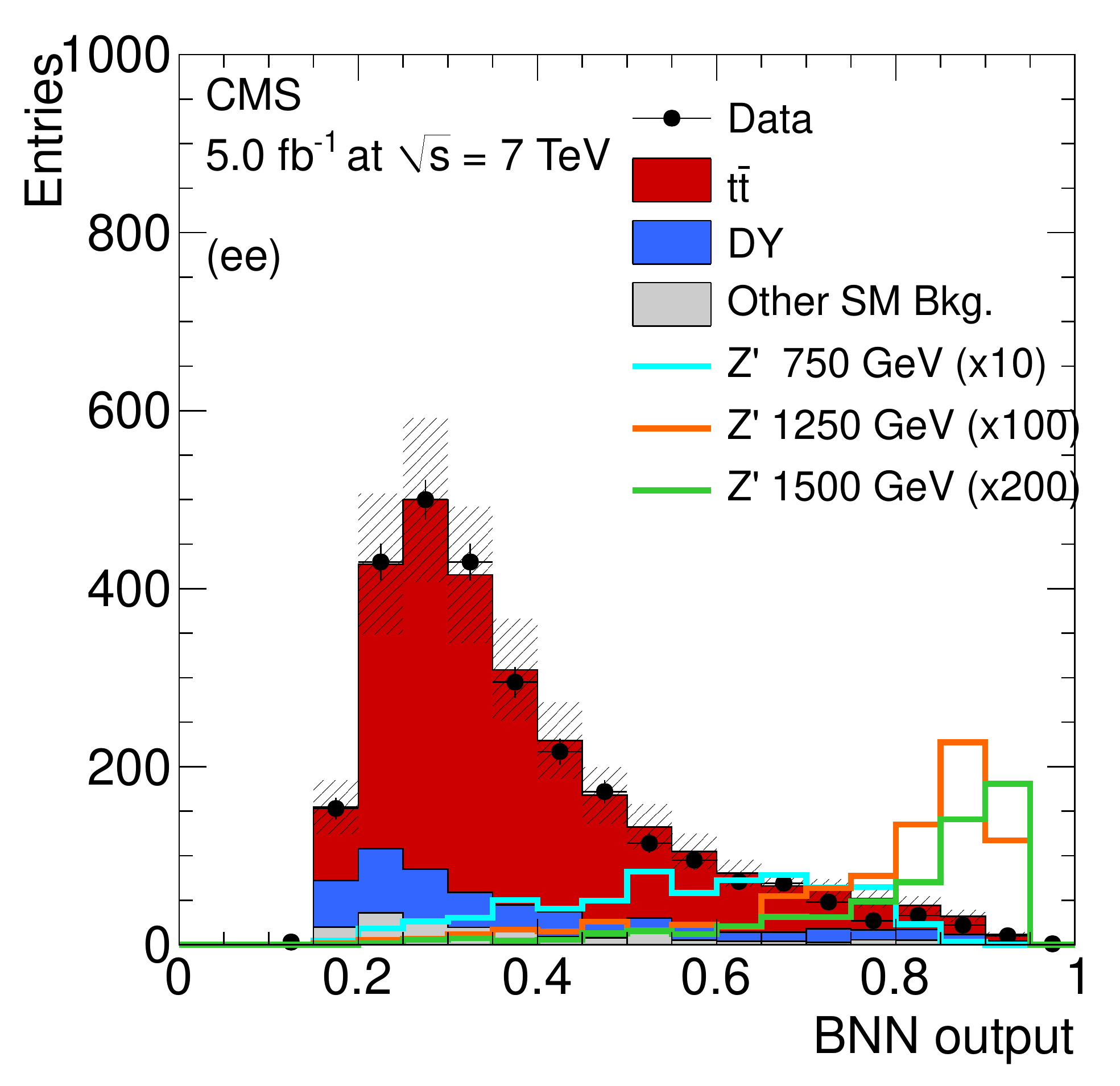}
  \includegraphics[width=\cmsFigWidth]{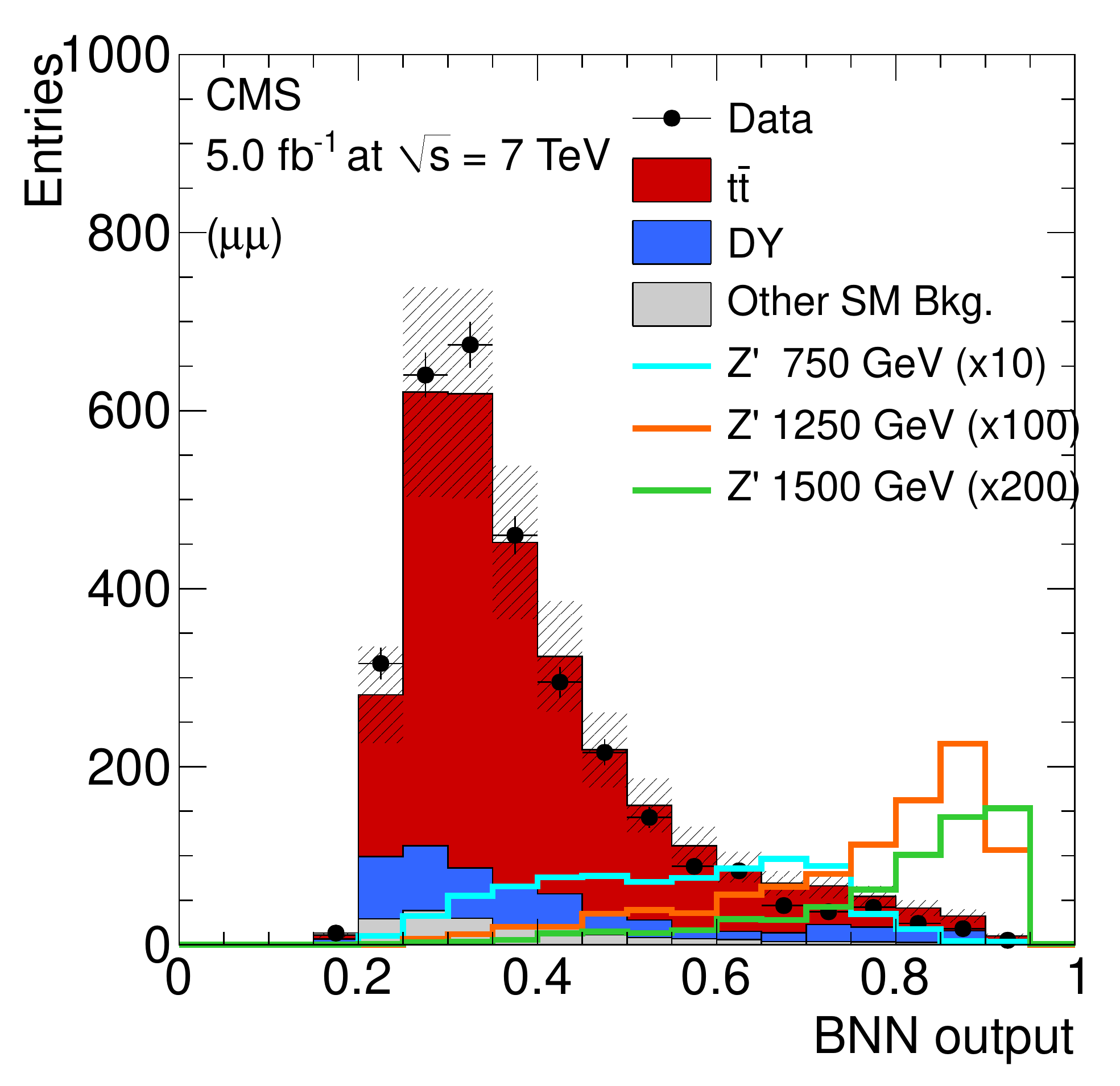}
  \includegraphics[width=\cmsFigWidth]{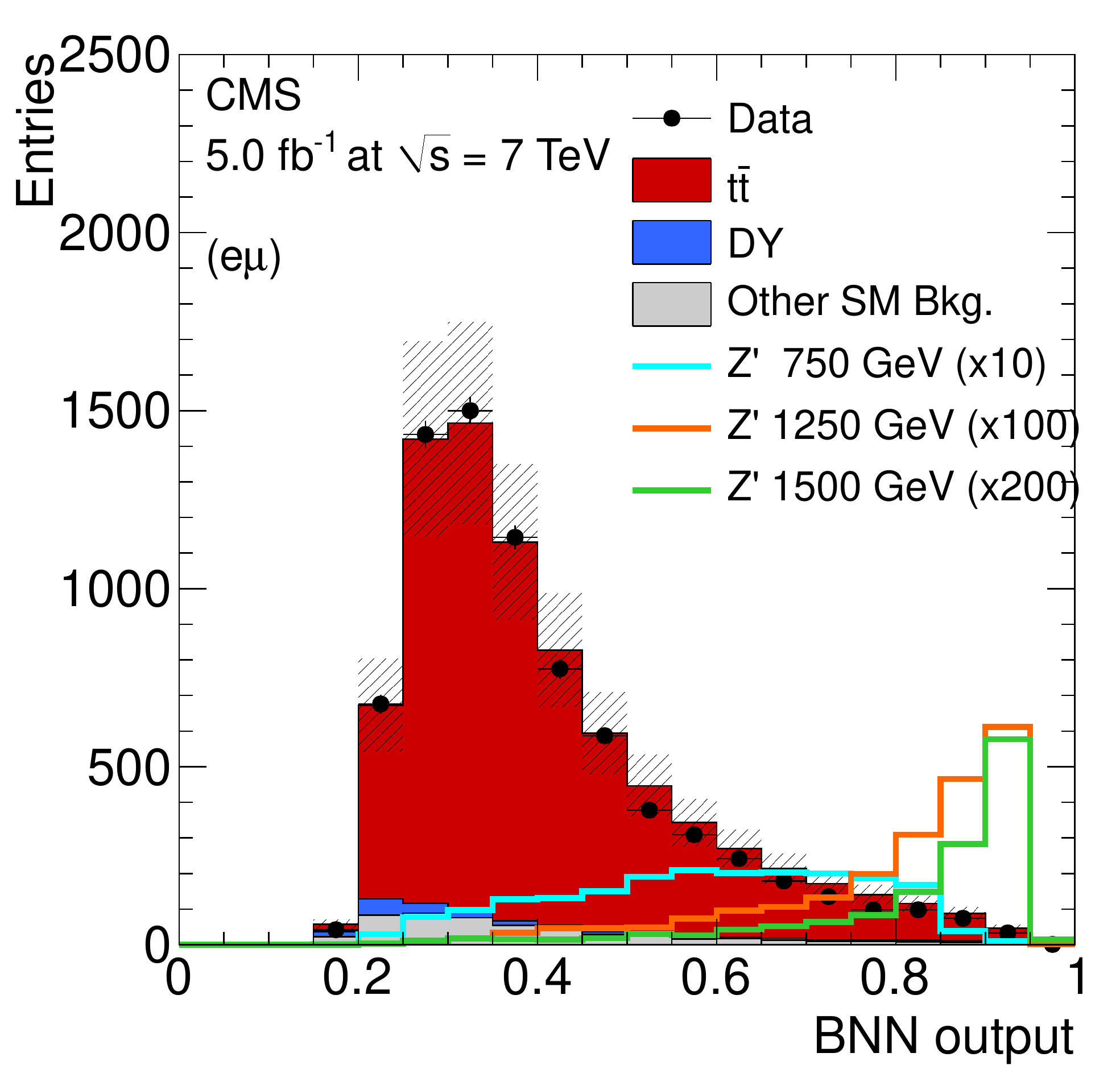}
 \caption{Distributions of the BNN output discriminant for the \Pe\Pe,
   $\mu\mu$, and $\Pe\mu$ channels. The
hatched region indicates systematic uncertainties on the sum of SM
backgrounds. The $\cPZpr$ signal corresponds to a resonance width of
$\Gamma_{\cPZpr} = 0.012M_{\cPZpr}$, and has been scaled up so as to be visible.
\label{fig:bnnout}}
\end{figure*}

\begin{figure*}[hbt]
 \centering
  \includegraphics[width=\cmsFigWidth]{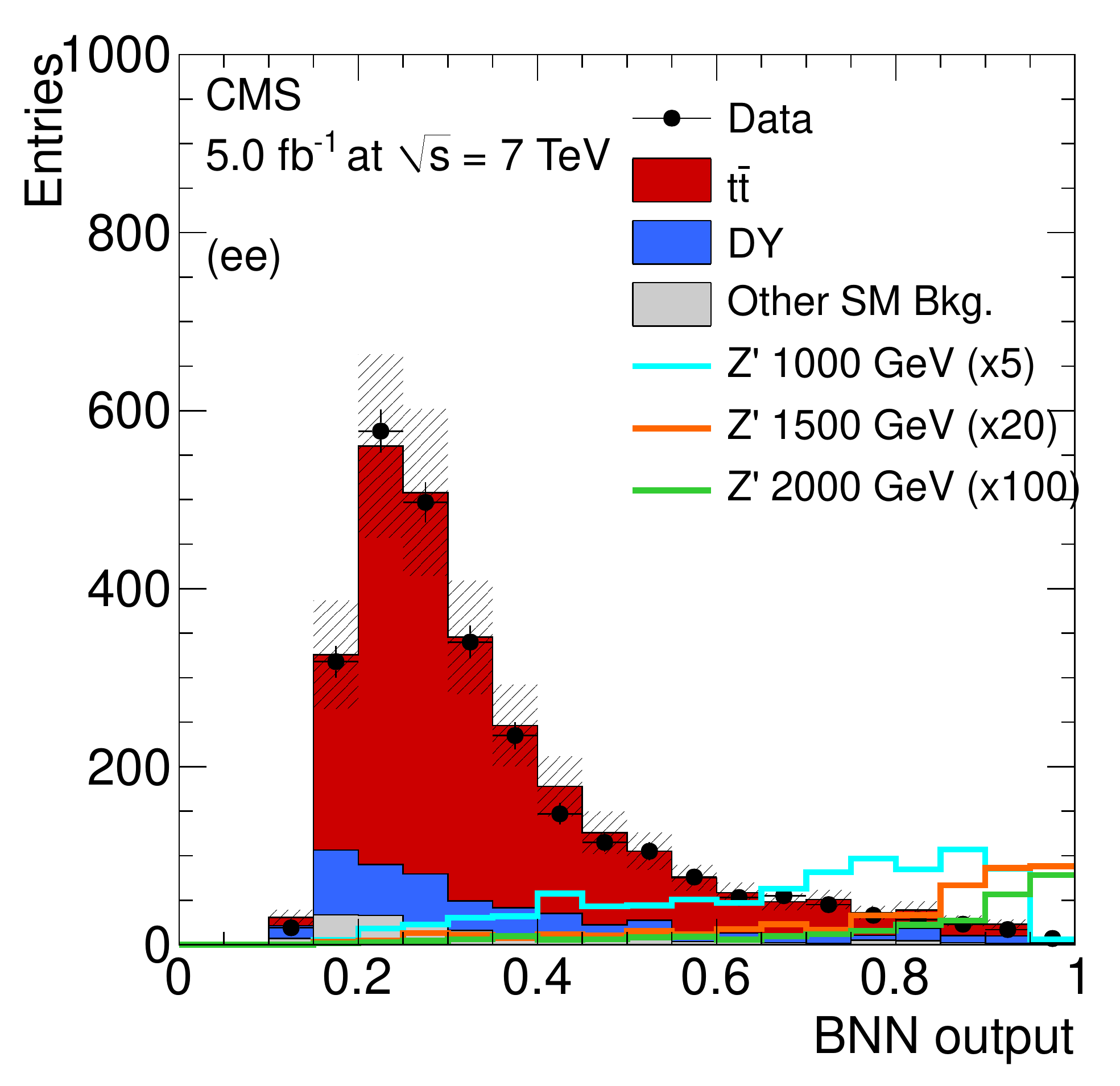}
  \includegraphics[width=\cmsFigWidth]{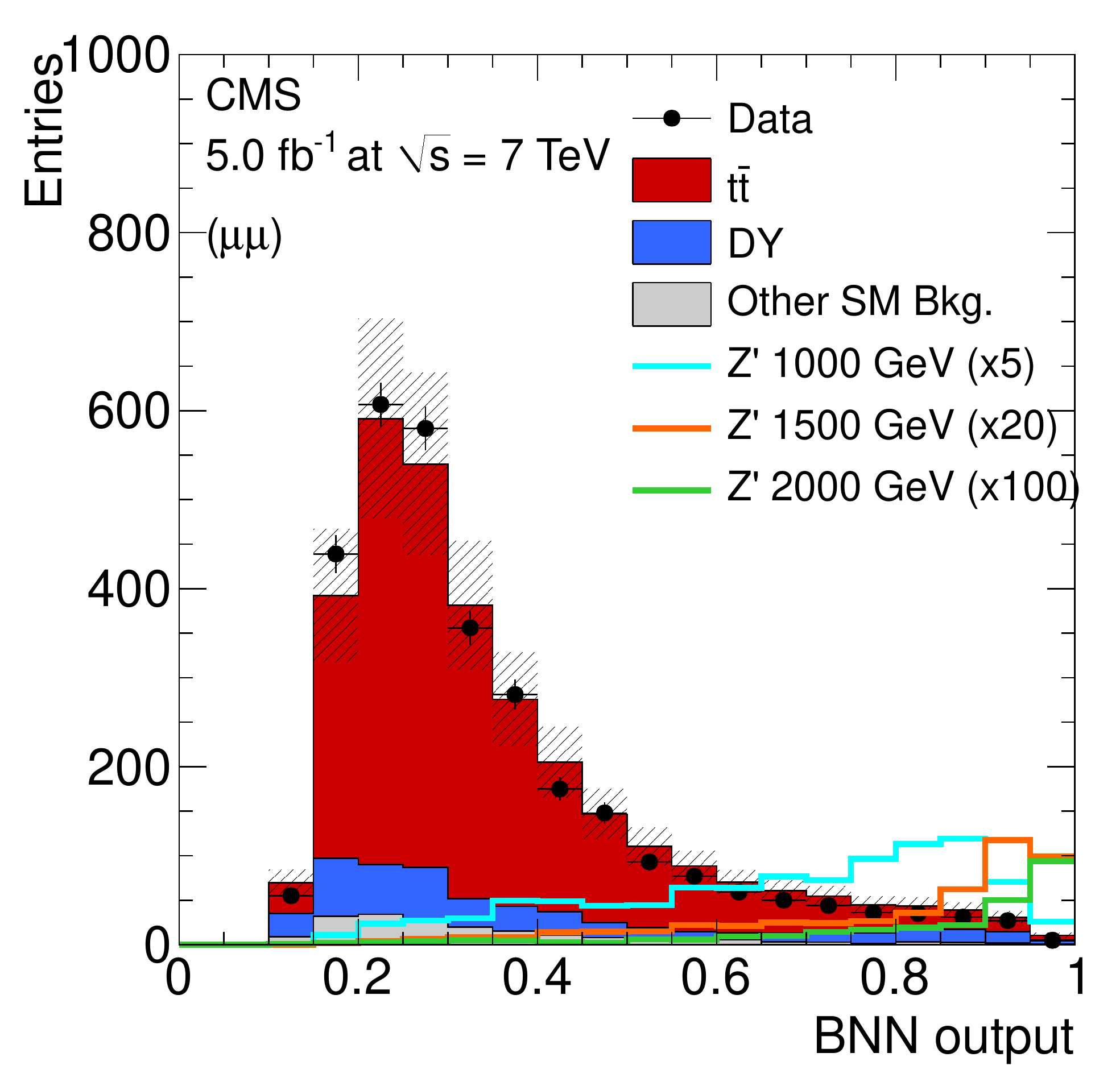}
  \includegraphics[width=\cmsFigWidth]{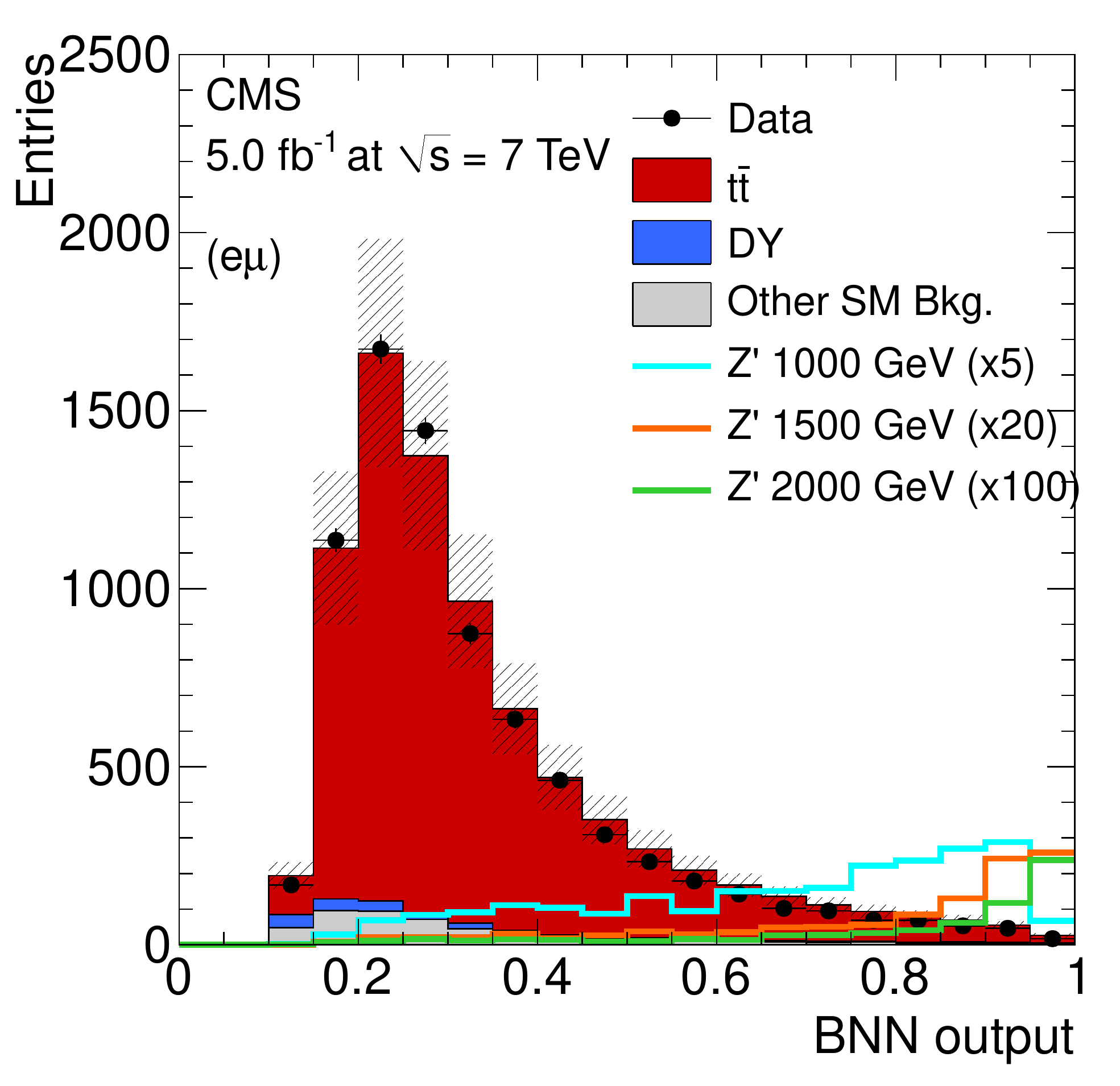}
 \caption{Distributions of the BNN output discriminant for the \Pe\Pe,
   $\mu\mu$, and $\Pe\mu$ channels. The
hatched region indicates systematic uncertainties on the sum of SM
backgrounds. The $\cPZpr$ signal corresponds to a resonance width of
$\Gamma_{\cPZpr} = 0.10M_{\cPZpr}$, and has been scaled up so as to be visible.
\label{fig:bnnout-Zprime10}}
\end{figure*}

The resulting BNN outputs for the observed data, the SM background, and
the $\cPZpr$ signals for $\Gamma_{\cPZpr} = 0.012M_{\cPZpr}$  and
$\Gamma_{\cPZpr} = 0.10M_{\cPZpr}$ are shown in Figs.~\ref{fig:bnnout}
and~\ref{fig:bnnout-Zprime10}, respectively, for the \Pe\Pe, $\mu\mu$, and
$\Pe\mu$ channels. There is good agreement between data and the SM background in all
three channels with no evidence of a resonance signal. Upper limits
are set on $\sigma_{\cPZpr}\mathcal{B}(\cPZpr\rightarrow\ttbar)$.

\section{Systematic uncertainties
\label{sec:systematics}}
The signal and background models are affected by a number of
systematic uncertainties, which are propagated into the limit
calculation. The uncertainties are divided into two categories:
those that affect only the overall normalization of a process
(``rate'') and those that affect also the distribution of the
BNN discriminant (``shape''). The rate effects include the uncertainty on
predicted cross section and normalization
for each SM background based on data, as discussed in
Section~\ref{sec:model}, and the uncertainties from integrated
luminosity, lepton identification and isolation, $\cPqb$-tagging
scale factor, jet energy scale, and pileup reweighting for both SM
background and $\cPZpr$ signals. Rate uncertainties are also
included for the SM $\ttbar$ and $\PW\rightarrow \ell\nu$ events from
variations in the renormalization and factorization scales ($\lambda$), and
the matching scale for jet production threshold between jets from
matrix-element generation in \MADGRAPH and parton showering
in \PYTHIA~\cite{mlm}. The nominal value of $\lambda$ is set to a
dynamical mass scale of
$(2M_\cPqt)^2~+~(\sum~\pt^\text{jet})^2$ for SM $\ttbar$ events, and
$(M_{\PW})^2~+~(\sum~\pt^\text{jet})^2$ with $M_{\PW}=80.4\GeV$
for $\PW\rightarrow \ell\nu$ events. The nominal value of matching
scale is set to 20\GeV. The $\lambda$ and
matching scales are each varied up and down
by a factor of two with respect to their nominal values in order to
estimate the uncertainty. The shape
effects include the change in shape of the BNN distributions from the
uncertainties from jet energy scale, pileup reweighting, and $\lambda$ and
matching scales. The uncertainty due to parton distribution functions
is negligible and therefore not included. The uncertainty on the
multijet background is dominated by the statistical uncertainty of
the same-sign samples in Eq.~\ref{eq:qcdisoeff}. All uncertainties are
summarized in Table~\ref{tab:syst}, where each row represents an
independent entity. The dominant source of systematic uncertainty in the
background estimate is due to the $\ttbar$ cross-section uncertainty of
15\% which is fully correlated between the channels. The total uncertainty
on the sum of all SM backgrounds is 18\%.
\begin{table}[!h!tbp]
\topcaption{Rate (top part) and shape (bottom part) uncertainties, from different components,
affecting the total expected number of signal and background events.
The uncertainties on the $\lambda$ and matching
scales in SM $\ttbar$, the jet energy scale, and the pileup reweighting
are indicated by their range across bins of the BNN distribution.
\label{tab:syst}}
\scotchrule[lc]
Component & Uncertainty [$\%$] \\\hline
$\ttbar$ cross section & 15 \\
DY normalization & 30 \\
Diboson cross section & 3.8 \\
$\PW\rightarrow \ell\nu$ cross section & 5.0 \\
Single top cross section & 7.7  \\
QCD normalization & 13 (\Pe\Pe), 18.2 ($\mu\mu$), 9.7 ($\Pe\mu$) \\
Integrated luminosity   & 2.2  \\
Lepton identification  & 2.0 \\
$\cPqb$-tagging scale factor & 10 \\
$\lambda$ scale ($\PW\rightarrow \ell\nu$) & 100  \\
Matching scale ($\PW\rightarrow \ell\nu$) & 100  \\
\hline
$\lambda$ scale ($\ttbar$) & 1.9--2.9 \\
Matching scale ($\ttbar$) & 3.4--5.5  \\
Jet energy scale & 0.3--4.7  \\
Pileup reweighting & 0.3--2.4  \\
\donescotchrule
\end{table}

\section{Results
\label{sec:results}}
\begin{figure*}[hbtp]
\centering
  \includegraphics[width=0.45\textwidth]{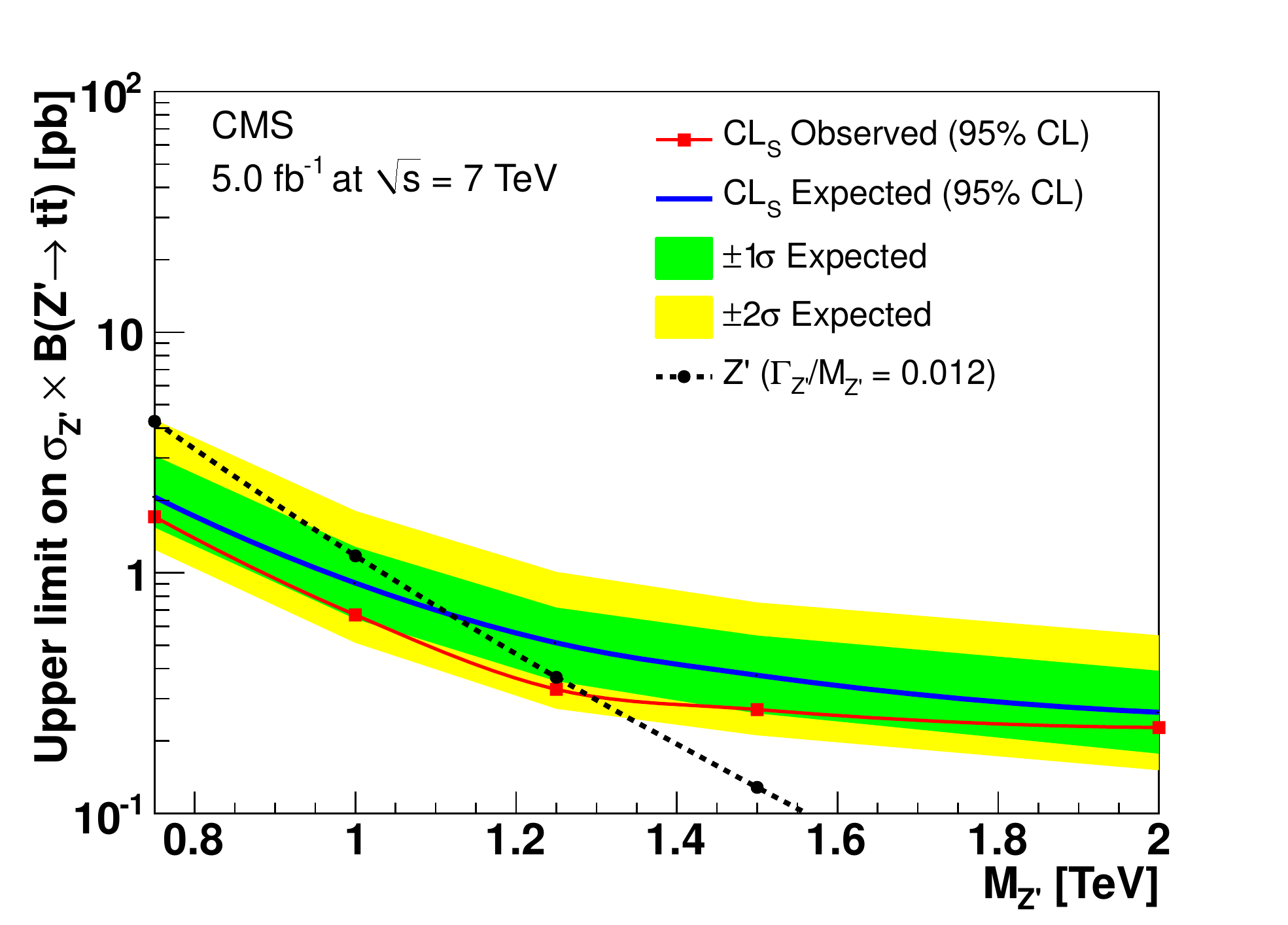}
  \includegraphics[width=0.45\textwidth]{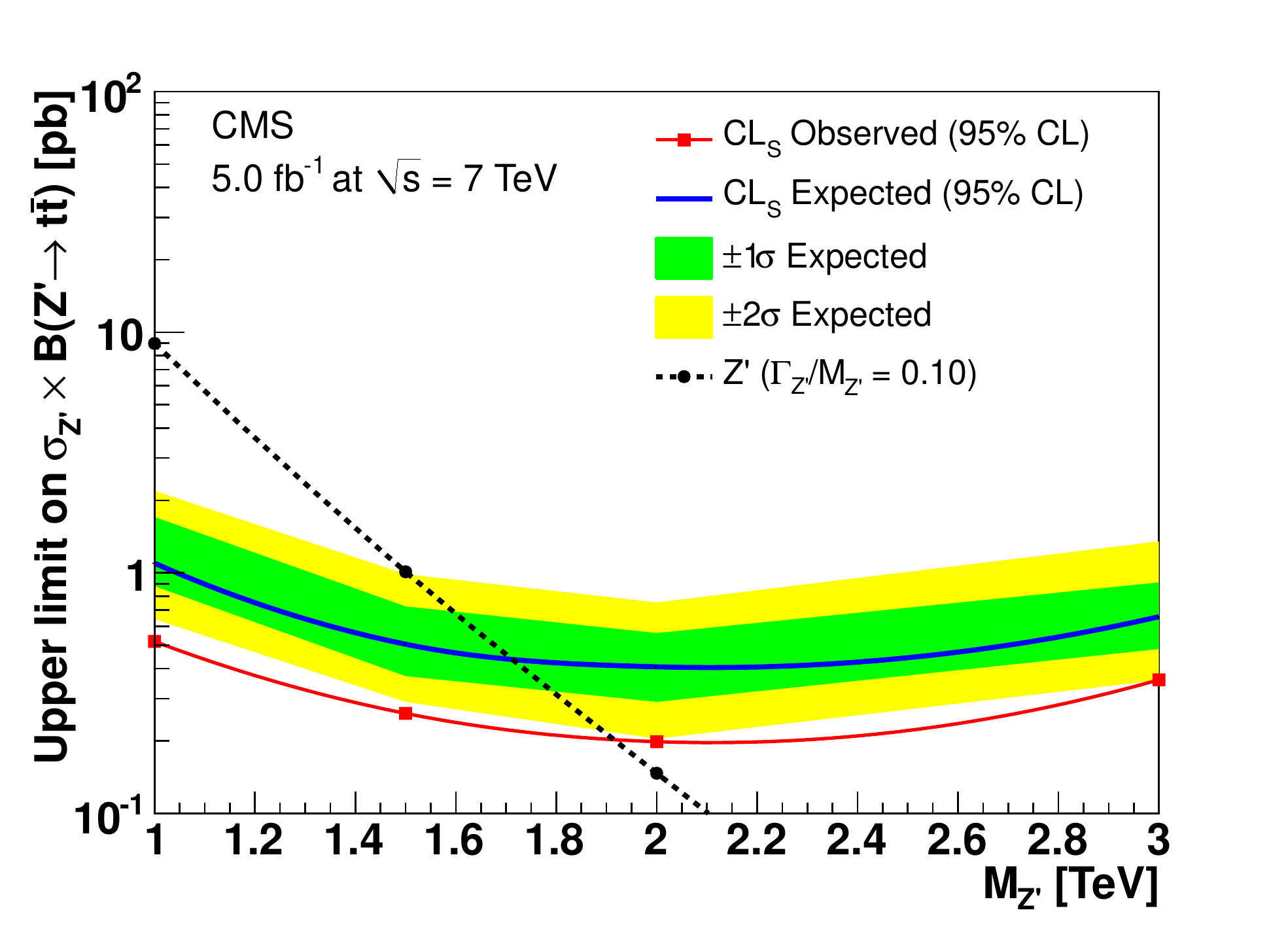}
 \caption{Upper limits on $\sigma_{\cPZpr}\mathcal{B}(\cPZpr\rightarrow\ttbar)$ versus $M_{\cPZpr}$ for
   narrow and wide resonances. Also shown is the theoretical prediction for the $\cPZpr$.
\label{fig:limit}}
\end{figure*}

With no excess observed, upper limits on
$\sigma_{\cPZpr}\mathcal{B}(\cPZpr\rightarrow\ttbar)$
at the $95\%$ confidence level (CL) for different values of
$M_{\cPZpr}$ are set using the \CLs
criteria~\cite{clsJunk,clsRead}.
All systematic effects are included with correlations across the different
samples and channels. The sensitivity of the results is estimated
using the invariant mass distributions shown in
Fig.~\ref{fig:invmass} and the BNN output distributions shown in
Fig.~\ref{fig:bnnout}, and comparing the expected limit on
$\sigma_{\cPZpr}\mathcal{B}(\cPZpr\rightarrow\ttbar)$ for the two
methods at $M_{\cPZpr} = 750$\GeV and
$\Gamma_{\cPZpr} = 0.012M_{\cPZpr}$. An expected limit is obtained from an ensemble
of simulated pseudo-data sets, where each set is constructed from the
background-only hypothesis. Using the BNN distribution improves the expected limit by
29\% compared to using invariant mass distribution. For this reason, the more
sensitive BNN technique is used for the subsequent measurements.
The resulting expected limits and the observed limits using data are shown in
Fig.~\ref{fig:limit} for both narrow and wide resonances.

The theoretical predictions for a
leptophobic $\cPZpr$~\cite{jainharris,ZprimeKfactor}
are used to exclude heavy $\cPZpr$ resonances of masses $M_{\cPZpr} < 1.3$\TeV
for a width $\Gamma_{\cPZpr} = 0.012M_{\cPZpr}$, and $M_{\cPZpr} < 1.9$\TeV
for a width $\Gamma_{\cPZpr} = 0.10M_{\cPZpr}$.  In the current analysis, the
expected lower limits on $M_{\cPZpr}$ are 1.1\TeV and 1.7\TeV
for $\Gamma_{\cPZpr} = 0.012M_{\cPZpr}$ and $\Gamma_{\cPZpr} = 0.10M_{\cPZpr}$,
respectively.

\section{Summary
\label{sec:summary}}
A data sample, corresponding to an integrated luminosity of
5.0\fbinv collected in $\Pp\Pp$ collisions at
$\sqrt{s} = 7$\TeV, has been analyzed in a search for heavy resonances
decaying to top quark-antiquark pairs with subsequent leptonic decay
of both top quark and antiquark. No excess beyond the standard model
prediction is observed. Upper limits at the 95\% CL are derived on the
product of the production
cross section and branching fraction for these decays, for various masses
of narrow and wide resonances. The existence of a leptophobic $\cPZpr$ topcolor
particle is excluded for $M_{\cPZpr} < 1.3$\TeV with $\Gamma_{\cPZpr} =
0.012M_{\cPZpr}$, and for $M_{\cPZpr} < 1.9$\TeV with $\Gamma_{\cPZpr} = 0.10M_{\cPZpr}$.

\section*{Acknowledgements}
We congratulate our colleagues in the CERN accelerator departments for
the excellent performance of the LHC and thank the technical and
administrative staffs at CERN and at other CMS institutes for their
contributions to the success of the CMS effort. In addition, we
gratefully acknowledge the computing centres and personnel of the
Worldwide LHC Computing Grid for delivering so effectively the
computing infrastructure essential to our analyses. Finally, we
acknowledge the enduring support for the construction and operation of
the LHC and the CMS detector provided by the following funding
agencies: BMWF and FWF (Austria); FNRS and FWO (Belgium); CNPq, CAPES,
FAPERJ, and FAPESP (Brazil); MEYS (Bulgaria); CERN; CAS, MoST, and
NSFC (China); COLCIENCIAS (Colombia); MSES (Croatia); RPF (Cyprus);
MoER, SF0690030s09 and ERDF (Estonia); Academy of Finland, MEC, and
HIP (Finland); CEA and CNRS/IN2P3 (France); BMBF, DFG, and HGF
(Germany); GSRT (Greece); OTKA and NKTH (Hungary); DAE and DST
(India); IPM (Iran); SFI (Ireland); INFN (Italy); NRF and WCU (Korea);
LAS (Lithuania); CINVESTAV, CONACYT, SEP, and UASLP-FAI (Mexico); MSI
(New Zealand); PAEC (Pakistan); MSHE and NSC (Poland); FCT (Portugal);
JINR (Armenia, Belarus, Georgia, Ukraine, Uzbekistan); MON, RosAtom,
RAS and RFBR (Russia); MSTD (Serbia); SEIDI and CPAN (Spain); Swiss
Funding Agencies (Switzerland); NSC (Taipei); ThEP, IPST and NECTEC
(Thailand); TUBITAK and TAEK (Turkey); NASU (Ukraine); STFC (United
Kingdom); DOE and NSF (USA).

\bibliography{auto_generated}   
\cleardoublepage \appendix\section{The CMS Collaboration \label{app:collab}}\begin{sloppypar}\hyphenpenalty=5000\widowpenalty=500\clubpenalty=5000\textbf{Yerevan Physics Institute,  Yerevan,  Armenia}\\*[0pt]
S.~Chatrchyan, V.~Khachatryan, A.M.~Sirunyan, A.~Tumasyan
\vskip\cmsinstskip
\textbf{Institut f\"{u}r Hochenergiephysik der OeAW,  Wien,  Austria}\\*[0pt]
W.~Adam, E.~Aguilo, T.~Bergauer, M.~Dragicevic, J.~Er\"{o}, C.~Fabjan\cmsAuthorMark{1}, M.~Friedl, R.~Fr\"{u}hwirth\cmsAuthorMark{1}, V.M.~Ghete, N.~H\"{o}rmann, J.~Hrubec, M.~Jeitler\cmsAuthorMark{1}, W.~Kiesenhofer, V.~Kn\"{u}nz, M.~Krammer\cmsAuthorMark{1}, I.~Kr\"{a}tschmer, D.~Liko, I.~Mikulec, M.~Pernicka$^{\textrm{\dag}}$, D.~Rabady\cmsAuthorMark{2}, B.~Rahbaran, C.~Rohringer, H.~Rohringer, R.~Sch\"{o}fbeck, J.~Strauss, A.~Taurok, W.~Waltenberger, C.-E.~Wulz\cmsAuthorMark{1}
% ...
\vskip\cmsinstskip
\dag:~Deceased\\
1:~~Also at Vienna University of Technology, Vienna, Austria\\
2:~~Also at CERN, European Organization for Nuclear Research, Geneva, Switzerland\\
% ...

\end{sloppypar}
\end{document}